\documentclass[a4paper,11pt]{article}
\usepackage{jheppub} % for details on the use of the package, please see the JINST-author-manual
\usepackage{lineno}
\usepackage{wrapfig}
\usepackage{caption}
\usepackage{subcaption}

\newcommand{\lr}[1]{\left( #1 \right)}
\newcommand{\lrt}[1]{\left[ #1 \right]}
\newcommand{\ltb}[1]{\Bigg[ #1 \Bigg]}

\newcommand{\lrs}[1]{\left\{ #1 \right\}}
\newcommand{\tx}[1]{\text{#1}}

\title{\boldmath Circular strings, magnons, plane waves and local quenches  in BTZ}

\author[a]{Justin R. David}
\affiliation[a]{Centre for High Energy Physics,
Indian Institute of Science,
C. V. Raman Avenue, Bangalore 560012, India.}
\emailAdd{justin@iisc.ac.in}
\author[b]{Rahul Metya}
\affiliation[b]{International Centre for Theoretical Sciences-TIFR,
Shivakote, Heseraghatta Hobli,
Bangalore North 560089, India.}
\emailAdd{rahul.metya@icts.res.in}

\abstract{We show that string theory on the geometry $BTZ\times S^3\times M$ supported with either Neveu-Schwarz flux or Ramond flux admits states which obey identical dispersion relations to those of classical solutions like circular strings, giant magnons,  or plane wave excitations in the geometry $ AdS_3 \times S^3 \times  M$. Here, $M$ can be $T^4$, $K3$, or $S^3\times S^1$. This is made possible by the map, which takes the particle at the origin of $AdS_3$ with angular momentum along one of the angles of $S_3$ to a particle 
falling into the BTZ horizon. We use this map to construct circular strings, magnons, as well as plane waves in the BTZ geometry. We show that the $SL(2, R)$ charges of these states on $AdS_3$ and that of the corresponding states in the BTZ geometry are related by a boost. The dual description of these states in the BTZ geometry are local quenches in the thermal CFT.  These quenches carry energy density, $R$-charges,  non-trivial expectation value of the marginal operator dual to the dilaton
 and move on the light cone in  CFT. In general, the left and the right moving quenches
 are not symmetric.}

\begin{document}
\maketitle
\flushbottom

\section{Introduction}
\label{sec:intro}

 The BTZ black hole \cite{Banados:1992wn,Banados:1992gq},
 It is one of the most well-studied solutions among black hole geometries. It also occurs as the near-horizon geometry of the $D1D5$ black hole in string theory \cite{Maldacena:1998bw}.
 It serves as a prototype for studies of black hole entropy, information paradox, and the holographic dual of large $c$ 
 CFTs at finite temperature. This is partly 
  because many aspects of the BTZ black hole are exactly solvable. Solutions for 
 geodesics,  wave equations of arbitrary spin fields \cite{Datta:2011za, Datta:2012gc} can be obtained 
 in closed form as it is locally $AdS_3$. 
 String propagation has also been studied earlier using the WZW model of the BTZ geometry.
 \cite{Natsuume:1996ij,Satoh:1997xe,Maldacena:2000hw,Maldacena:2000kv,Hemming:2001we,Hemming:2002kd}.
 However, a clear physical interpretation of the states of the string is lacking. This is important due to the existence of the horizon. Clearly, there must be string states that fall into the horizon, and these have not been identified and studied in earlier work.
 
  In this paper, we identify string states that fall into the horizon of the BTZ black hole. The background we consider is string theory on $BTZ \times S^3 \times M$, where $M$ can be $T^4, K3$, or $S^3 \times S^1$. A class of states that play an important role in discussions of integrability in the $AdS$
   backgrounds are giant magnons, circular strings, and the plane wave spectra. The construction of all these states for the $AdS_3$ background begins with a geodesic at the origin of $AdS_3$. For instance, the plane wave spectra are obtained by zooming into the geometry around the BPS geodesic at the origin of $AdS_3$ with angular momentum along one of the angles of the $S^3$ \cite{Berenstein:2002jq}. 
   This geodesic is light-like when considered on $AdS_3 \times S^3$, but time-like
when considered only on $AdS_3$. There exists a map that relates the time-like geodesic at the origin in $AdS_3$ to an in-falling geodesic in BTZ \cite{Casini:2011kv}.  To be explicit, this map is given by the following relations
    \begin{align}\label{coordtrans}
        &X_0 =\sqrt{r^2+R^2}\ \sin{t} =\frac{R}{M} \sqrt{\rho^2-M^2} \sinh{M\tau},  \nonumber\\
        &X_1=\sqrt{r^2+R^2} \cos{t} =\frac{R}{M}\lrt{\sqrt{\rho^2-M^2} \sinh\eta_1 \cosh{M\tau} + \rho \cosh{\eta_1} \cosh{Mx}},  \nonumber\\
        &X_2=r\sin{\phi}=\frac{R}{M} \rho \sinh{Mx}, \\
    &X_3= r \cos{\phi}= \frac{R}{M}\lrt{\sqrt{\rho^2-M^2} \cosh{\eta_1} \cosh{M\tau} + \rho \sinh{\eta_1} \cosh{Mx}} .\nonumber
\end{align}
Note that the embedding coordinates satisfy 
\begin{eqnarray}
-X_0^2 - X_1^2 + X_2^2 + X_3^3 =- R^2, 
\end{eqnarray}
where $R$ is the radius of $AdS_3$. The induced metric in the $t, r, \phi$ coordinates is global $AdS_3$, and 
in the $\tau, \rho, x$ coordinates is BTZ, $M$ is the location of the horizon. It can be seen that this map relates the particle at the origin 
$r=0$ to that of a geodesic in-falling into the BTZ horizon. This map depends on the parameter $\eta_1$, which can be related to the position of the particle when it is released in the BTZ geometry. We use this map and its generalisation, which allows us to give velocity along the $x$ direction to construct and study giant magnons and circular strings in BTZ. 

The classical solutions in the WZW  sigma model of the BTZ background carry conserved 
$SL(2, R)$ charges. 
We evaluate these charges and show that they are related to the charges of the corresponding solution 
in $AdS_3$ by a $SO(1, 2)$ boost. 
This boost parameter  depends on  the map  relating geodesics in $AdS_3$ and BTZ
 and for the map given in (\ref{coordtrans}) it turns out to be $ \eta_1$.   The resulting solutions in BTZ obey the same dispersion relation as that of the solution in $AdS_3$. This is because the dispersion relation results from the Virasoro condition, which relates the $SL(2, R)$ and the $SU(2)$ Casimirs. The $SL(2, R)$ Casimir remains invariant under the $SO(1, 2)$ boosts and therefore the Virasoro condition and consequently the dispersion relation is BTZ and $AdS_3$ are identical.

We then study the plane wave spectrum in the BTZ geometry. 
Recall in $AdS_3 \times S^3$, 
The plane wave limit and its spectrum of string states are obtained by zooming into the geometry near a lightlike geodesic. This particle sits at the origin in $AdS_3$ and spins along one of the isometries of $S^3$. We use the map (\ref{coordtrans}) and its generalisation, which relates the time-like geodesic at the origin in $AdS_3$ to an in-falling geodesic in BTZ with angular momentum along one of the circles of $S^3$, to zoom into this geometry of the BTZ.  Plane wave limits of in-falling geodesics into black hole horizons have been studied earlier, see for example in \cite{PandoZayas:2002dso}, the resulting metric depends on the light cone co-ordinate. 
The limit we obtain by zooming in near in-falling geodesics to the BTZ horizon with angular momentum 
on $S^3$  results in a homogeneous plane wave metric. In fact, 
the  plane wave dispersion relation in the BTZ geometry is identical to that of the $AdS_3$ case, which reflects the local  
$AdS_3$ nature of the BTZ geometry, and it does not depend on the parameters of the map.

Finally, we examine the interpretation of the states that we have found in the BTZ geometry. 
These states should be excitations of the thermal state of the corresponding boundary CFT. 
 The boundary CFT in this case
is the symmetric product CFT on $T^4$ or $K3$, or in the case $S^3 \times  S^1$, it is the instanton moduli space on $S^3 \times S^1$. 
 We show that the classical states in the bulk, geodesics, circular strings, and magnons are dual to 
 local quenches similar to that studied in \cite{Caputa:2014eta} for the 3 dimensional BTZ geometry 
 which carry only energy density and move on the light cone in the CFT.
 Here, however, the quenches in addition to energy density 
 carry  $R$ charges which arise due to the angular momentum on $S^3$.   These states also 
 have a non-trivial expectation value of the marginal operator dual to the dilaton in the CFT. 
  We show that for the more general case of
 the map in (\ref{coordtrans}), which allows longitudinal velocity of the in-falling particle in the $x$ direction, the dual quenches on the 
 left and right light cones are not symmetric, and the amplitude of the pulse differs. Thus, these quenches generalise those found first in  \cite{Caputa:2014eta}. 
 The figure \ref{AsymmetricQuenchStressTensor} shows the plot of the energy density of these quenches, which 
 clearly shows the asymmetric left and right-moving pulses travelling at the speed of light. 
 
 The organisation of this paper is as follows. 
 In section \ref{sec2}, we study the map which relates the time-like geodesic of a particle at the origin of global $AdS$ to that 
 of an in-falling particle in BTZ in detail, and its generalisation. By examining the WZW sigma model on global $AdS$ and that of $BTZ$, we show that the conserved charges of the geodesics in each of the spaces are related by an $SO(1, 2)$ transformation. In section \ref{sec3}, we zoom into the geometry of the in-falling geodesic in BTZ and show that the geometry reduces to that of a homogeneous plane with identical to that in $AdS_3$.   In section \ref{sec4}, we derive the description of the in-falling classical solutions in the BTZ geometry as local quenches in the thermal 2-dimensional CFT. Section \ref{sec5} has our conclusions.  
 Appendix  \ref{A:penrose limit of AdS5S5} reviews the Penrose limits of the Schwarzschild black hole in $AdS_5\times S^5$.

\section{Geodesics  and their charges }
\label{sec2}

$AdS_3$ and BTZ spacetime are locally isomorphic. In fact, the metric of both the space times 
can be constructed by different parametrisations of the hyperboloid
\begin{eqnarray}
-X_0^2  -X_1^2 + X_2^2 + X_3^2 = -R^2.
\end{eqnarray}
Let us recall how this comes about, consider the embeddings 
\begin{eqnarray} \label{embedding}
X_0 &=& \sqrt{r^2 + R^2 } \sin t =  \frac{R}{M} \sqrt{\rho^2 - M^2} \sinh M \tau, \\ \nonumber
X_1 &=& \sqrt{r^2 + R^2} \cos t  = \frac{R}{M} \rho  \cosh M x , \\ \nonumber
X_2 &=& r\sin\phi  = \frac{R}{M}  \rho \sinh M x , \\ \nonumber
X_3 &=& r\cos \phi = \frac{R}{M} \sqrt{\rho^2 - M^2} \cosh M \tau. 
\end{eqnarray}
The induced metric  on the hyperboloid in the $t, r, \phi$ coordinates is that of $AdS_3$
\begin{eqnarray} \label{ads3}
ds^2_{AdS_3}  = -( r^2 + R^2) dt^2 + \frac{R^2}{ r^2 + R^2} dr^2 + r^2 d\phi^2 . 
\end{eqnarray}
It is understood that we need to think of $t$ as the coordinate on the covering space in global $AdS_3$. 
In the  $\tau, \rho, x$ coordinates,  the metric is given by 
\begin{eqnarray} \label{btzmetric}
ds^2_{\rm BTZ} = R^2 \left[ - (  \rho^2 - M^2) d\tau^2 + \frac{ d\rho^2}{ \rho^2 - M^2}  + \rho^2 dx^2 \right]. 
\end{eqnarray} 
The radius of curvature of both the spaces is set by $R$, while the location of the horizon in BTZ is given by 
$M$. Note that we obtain the Rindler BTZ metric with a planar horizon, identifying $ R x\sim  R x + 2\pi $ results in the BTZ black hole with a circular horizon. We will work with the Rindler BTZ in this paper. The map between $AdS_3$ and $BTZ$  coordinates given in (\ref{embedding}) allows us to relate geodesics in these spaces. We will exploit this fact to construct classical string solutions in $BTZ$ as well as to identify the geodesics we need to zoom in to obtain the plane wave limit. We can choose to embed  BTZ in the hyperboloid differently. For example, we can introduce a boost between the parametrisation of $X_1$ and $X_3$ in the BTZ embedding  (\ref{embedding}), and still obtain the metric of the BTZ.  With this boost, the map becomes 
\subsection{The in-falling particle in BTZ}
Consider the  following coordinate transformation 
    \begin{align}\label{coordinate transformation for in-falling geodesic}
        &X_0 =\sqrt{r^2+R^2}\ \sin{t} =\frac{R}{M} \sqrt{\rho^2-M^2} \sinh{M\tau}, \nonumber\\
        &X_1=\sqrt{r^2+R^2} \cos{t} =\frac{R}{M}\lrt{- \sqrt{\rho^2-M^2} \sinh\eta_1 \cosh{M\tau} + \rho \cosh{\eta_1} \cosh{Mx}}, \nonumber\\ \nonumber
        &X_2=r\sin{\phi}=\frac{R}{M} \rho \sinh{Mx}, \\ 
    &X_3= r \cos{\phi}= \frac{R}{M}\lrt{\sqrt{\rho^2-M^2} \cosh{\eta_1} \cosh{M\tau} - \rho \sinh{\eta_1} \cosh{Mx}} .
\end{align}
Note that the induced metric in the $\tau, \rho, x$ coordinate still remains BTZ.  Given the map  
(\ref{coordinate transformation for in-falling geodesic}) between $AdS_3$ and BTZ,  let us examine the trajectory of a time-like geodesic at the origin of $AdS_3$  in BTZ. The geodesic in $AdS_3$ is given by 
\begin{eqnarray} \label{origeod}
t= \kappa \, \sigma^0,  \qquad r(\sigma^0) = 0, \qquad \phi(\sigma^0) = 0 .
\end{eqnarray}
From the map in (\ref{coordinate transformation for in-falling geodesic}), we see that the BTZ co-ordinates on this trajectory must satisfy
\begin{align} \label{eq1}
  \sqrt{\rho^2-M^2} \cosh{\eta_1}\cosh{M\tau}- \rho \sinh{\eta_1}=0  , \qquad  x = 0.
\end{align}
Here, it is understood that $\tau, \rho$ are functions of the affine parameter $\sigma^0$. Now, from the  first  2 equations  (\ref{coordinate transformation for in-falling geodesic}), we obtain  the relations 
\begin{eqnarray} \label{eq2}
R \sin t &=& \frac{R}{M} \sqrt{ \rho^2 - M^2} \sinh M \tau, \\ \nonumber
R \cos t &=& \frac{R}{M} \frac{\rho}{\cosh \eta_1} .
\end{eqnarray}
Here we have used $r=0$ and the relation  (\ref{eq1}) to obtain the second line. Differentiating (\ref{eq1}) with respect to $\sigma^0$, we obtain 
\begin{eqnarray}\label{eq3}
   \dot \tau  \cosh \eta_1 \sinh M \tau=-\frac{M \dot \rho}{  ( \rho^2 - M^2)^{\frac{3}{2} }} \sinh \eta_1.
\end{eqnarray}
Similarly differentiating the first equation of (\ref{eq2}), we obtain
\begin{eqnarray}
R \kappa \cos t &=& \frac{R}{M} \left( \frac{ \rho \dot \rho}{\sqrt{ \rho^2 -M^2}} \sinh M \tau  + M \dot\tau \sqrt{\rho^2- M^2} \cosh M\tau \right ), \\ \nonumber
&=& \frac{R}{M} \frac{\rho \kappa }{ \cosh \eta_1} .
\end{eqnarray}
Here we have used the trajectory $t= \kappa \sigma^0$ in $AdS_3$. Now using (\ref{eq3}) and (\ref{eq1})  to eliminate $\dot \rho$ and  $\tau$, we obtain the 
relation 
\begin{eqnarray} \label{geoeq1}
 (\rho^2 - M^2) \dot \tau = \kappa M  \sinh \eta_1 .
\end{eqnarray}
Substituting this relation and eliminating the dependence on $\tau$ in (\ref{eq3}), we obtain 
\begin{eqnarray} \label{geoeq2}
 - \frac{M^2 \kappa^2 \sinh^2 \eta_1 }{ \rho^2 - M^2} + \frac{\dot \rho^2}{\rho^2 - M^2}  =-  \kappa^2 .
\end{eqnarray}

\subsubsection*{Comparison with the in-falling geodesic in BTZ}

Consider time like geodesic equations in the BTZ metric given in (\ref{btzmetric}).  The isometry along the $\tau$ and $x$ direction leads to the following conserved quantities
\begin{eqnarray} \label{constants}
-R^2 (\rho^2 -M^2) \dot \tau = E , \qquad  R^2 \rho^2 \dot x = v .
\end{eqnarray}
The condition that the geodesic is time-like with mass $\tilde m $ leads to the equation
\begin{eqnarray}
-R^2 ( \rho^2 - M^2) \dot \tau^2 + R^2 \frac{ \dot \rho^2 }{ \rho^2 - M^2}  + R^2 \rho^2 \dot x^2  = - \tilde m^2 .
\end{eqnarray}
We can eliminate $\dot \tau, \dot x$ using the constants of motion (\ref{constants}) and obtain 
\begin{eqnarray} \label{constants3}
- \frac{E^2}{ R^4 ( \rho^2 - M^2) } + \frac{\dot \rho^2}{ \rho^2 - M^2} + \frac{v^2}{  R^4 \rho^2 } = 
- \frac{\tilde m^2}{R^2} .
\end{eqnarray}
Comparison of these equations with the equation of the trajectory obtained in (\ref{geoeq1}) and (\ref{geoeq2}), we can identify it as the in-falling geodesic with  the following constants of motion
\begin{eqnarray} \label{constants1}
\frac{E}{R^2} = - \kappa M \sinh \eta_1,   \qquad v =0, \qquad  \frac{\tilde m ^2}{R^2 } =  \kappa^2 .
\end{eqnarray}
Once this map is made, the boost parameter $\eta_1$ can be related to the point of the release of the particle from (\ref{geoeq2}).  We see that $\dot \rho = 0$ at 
\begin{eqnarray} \label{constants2}
\rho|_{\dot \rho =0} = M \cosh \eta_1.
\end{eqnarray}
Therefore, we have concluded that the map (\ref{coordinate transformation for in-falling geodesic}) takes the time-like geodesic at the centre of $AdS_3$   to 
a radially in-falling geodesic in BTZ released at radial distance $M \cosh \eta_1$. 
The initial conditions of this geodesic at $\sigma^0=0$ are 
\begin{eqnarray} \label{initialcond}
&& x(0) = 0 ,\;\; \dot x(0) =0, \qquad \qquad
\tau (0) = 0, \;\; \dot\tau(0)  =  \frac{ \kappa}{ M  \sinh\eta_1}, \\ \nonumber
&& \rho(0) = M \cosh\eta_1,\;\;  \dot\rho =0.
\end{eqnarray}
This can be seen by using the equations 
 (\ref{constants}), (\ref{constants1}), (\ref{constants2}) and (\ref{eq2}) at $\sigma^0 =0$. 
 
\subsubsection*{$SL(2, R) $ charges of the in-falling particle}

The geodesics both in $AdS_3$ and $BTZ$ are solutions to the string sigma model on $AdS_3\times S^3 \times M$ or $BTZ \times S^3 \times M$. 
Let us describe the bosonic part of the sigma model on $AdS_3$ or $BTZ$  by the  $SL(2, R)$ WZW action
\begin{eqnarray} \label{sl2rsigma}
S  &=& S_{SL(2, R) } + S_{SU(2)} + S_{M} , \\ \nonumber
  S_{SL(2, R)}  &=&-\frac{k}{8\pi}\int_\Sigma 
   d^2\sigma\ Tr\lrt{g^{-1}\partial_a g\ g^{-1}\partial^ag}  + \frac{ k}{12\pi}\ q \int_{\cal B}
   Tr\lrt{g^{-1}dg \wedge g^{-1}dg \wedge g^{-1}dg}  .
  \end{eqnarray}
Here $S_{SU(2)}$ refers to the $SU(2)$ WZW model on $S^3$ and $S_M$ refers to the sigma model on $M$. 
For the present, let us focus on the $SL(2, R) $ WZW model. 
The level $k$ is related to the radius of $AdS_3$ or $BTZ$ by $\alpha' k = R^2$. The parameter $q$ in front of the 
topological term lies between $0\leq q\leq 1$. When $q=0$, the background is supported by pure Ramond flux, and when $q=1$, the model contains pure Neveu-Schwarz flux.  The index $a$ runs over the world sheet co-ordinates  $\sigma^0, \sigma^1$ with the signature  $(-1, 1)$. The world sheet space coordinate  $\sigma^1$  runs from $0$ to $2\pi$ when describing closed strings. $\Sigma$ is the world sheet, which is the boundary of the  3-manifold ${\cal B}$. The $SL(2, R)$ model can be used to describe both $AdS_3$ and $BTZ$ by appropriate parametrisation of the group element $g$. In terms of the embedding coordinates, $g$ is given by 
\begin{eqnarray}
  g= \frac{1}{R} \begin{pmatrix}
      X_1 + X_2 & X_3+X_0\\
      X_3-X_0& X_1-X_2
  \end{pmatrix} ,  
  \quad\quad -X_0^2-X_1^2+X_2^2+X_3^2=- R^2. 
\end{eqnarray}
To describe string propagation on the global $AdS_3$, we choose the embedding as in \cite{Maldacena:2000hw}. 
\begin{eqnarray}
&& X_0 = \sqrt{r^2 + R^2} \, \sin t , \qquad X_1 = \sqrt{r^2 + R^2} \cos t , \\ \nonumber
&& X_2 = r \sin \phi, \qquad X_3 = r \cos \phi.
\end{eqnarray}
On the other hand, to describe string propagation in BTZ, we choose the embedding
\cite{Hemming:2001we}. 
\begin{eqnarray} \label{btzembedd}
&& X_0 = \frac{R}{M} \sqrt{\rho^2 - M^2} \sinh M \tau, 
\qquad X_1 = \frac{R}{M} \rho \cosh M x, \\ \nonumber
&& X_2 = \frac{R}{M} \rho \sinh M x , \qquad X_3 = \frac{R}{M} \sqrt{\rho^2 -M^2} \cosh M\tau.
\end{eqnarray}

The sigma model exhibits left and right $SL(2, R)$  conserved charges whose expressions are given by 
\begin{eqnarray} \label{defcharge1}
Q_L = \frac{1}{2\pi } \int_0^{2\pi} d\sigma^1 \Big( g^{-1}\dot g + q \,  g^{-1} g'  \Big) ,
\qquad 
Q_R =  \frac{1}{2\pi } \int_0^{2\pi } d\sigma^1 \Big(   \dot g  g^{-1}  - q \,  g' g^{-1} \Big) .
\end{eqnarray}
Here $\dot g$ refers to derivative with the world sheet time coordinate $\sigma^0$ and $g'$ refers to derivative with the world sheet space coordinate $\sigma^1$
Since $g \in SL(2, R)$, these charges belong  to the Lie algebra of $SL(2, R)$ and we can project them along its 
$3$ generators as follows
\begin{align}
   Q_{L,R}^{(i)} =i\ {\rm Tr} \lrt{\Sigma^i Q_{L,R}} ,\end{align}
where the $\Sigma$'s are related to the Pauli matrices $\hat \sigma_i$ as 
\begin{eqnarray}
\Sigma^1=\frac{i}{2}\hat \sigma_1, \qquad \Sigma^2=\frac{i}{2}\hat \sigma_3, \qquad \Sigma^3=-\frac{1}{2}\hat \sigma_2.
\end{eqnarray}

Let us evaluate the $SL(2, R) $ charges on the time like geodesic sitting at the origin described by the trajectory (\ref{origeod}). Note that for geodesics, the contribution proportional to $q$ in the charges (\ref{defcharge1}) vanishes. The left and right $SL(2, R)$ charges for this geodesic are given by 
 \begin{align} \label{oricharge}
     Q^{(1)}_{L,R}|_{{\tiny AdS_3}} =0,\quad Q^{(2)}_{L,R}|_{{\tiny AdS_3}}=0,\quad Q^{(3)}_{L,R}|_{{\tiny AdS_3} }=\kappa
 \end{align}
We proceed with the evaluation of the charges of the in-falling geodesic in BTZ  related to a time-like particle at the origin of $AdS_3$  by the map in (\ref{coordinate transformation for in-falling geodesic}). For this, let us substitute the embedding  (\ref{btzembedd}) into $g$ and evaluate the charges. For the  left $SL(2, R)$ charges,  we obtain
\begin{eqnarray} \label{defleftchg}
Q_L^{(1)}& =&   - \frac{  \dot \rho   }{\sqrt{ \rho^2 -M^2} }  \cosh M ( \tau - x)  - \frac{ ( \dot x + \dot \tau) }{M}
\rho \sqrt{ \rho^2 -M^2}  \sinh M ( \tau - x),   \\ \nonumber 
Q_L^{(2)} &=& \frac{ M^2 - \rho^2 }{M}  \dot \tau - \frac{\rho^2}{M} \dot x , \\ \nonumber
Q_L^{(3)}  &=&    \frac{  \dot \rho   }{\sqrt{ \rho^2 -M^2} }  \sinh M ( \tau - x)  + \frac{ ( \dot x + \dot \tau) }{M}
\rho \sqrt{ \rho^2 -M^2}  \cosh M ( \tau - x).
\end{eqnarray}
Since these charges are constants of motion, we can evaluate them by substituting the initial conditions  (\ref{initialcond}) for the in-falling particle. We find that the charges are given by 
\begin{eqnarray}
Q_L^{(1)}|_{\tiny BTZ}  = 0,  \qquad 
Q_L^{(2)} |_{\tiny BTZ} =  -\kappa \sinh\eta_1, \qquad Q_L^{(3)}|_{\tiny BTZ} =  \kappa \cosh\eta_1
\end{eqnarray}
Now, comparing the charges of the time like geodesic at the origin (\ref{oricharge})  to that of the in-falling particle in BTZ, we find that 
\begin{eqnarray} \label{boost1}
\left( \begin{array}{c}
Q_L^{(1)} \\ Q_L^{(2)} \\ Q_L^{(3)} 
\end{array}
\right)_{BTZ} = 
\left( \begin{array}{ccc}
1 & 0  & 0 \\
0 & \cosh\eta_1 &- \sinh\eta_1 \\
0 &- \sin\eta_1 & \cosh\eta_1 
\end{array}
\right) 
\left( \begin{array}{c}
Q_L^{(1)} \\ Q_L^{(2)} \\ Q_L^{(3)} 
\end{array}
\right)_{AdS_3}.
\end{eqnarray}
Therefore, the charges of the in-falling particle in BTZ, which is obtained from the time-like particle at the centre of $AdS_3$  by the map, are related by an $SO(1, 2)$ boost. There is another way to see that the charges of the geodesics of interest are related by boosts. 
Form the map in (\ref{coordinate transformation for in-falling geodesic}), We see that on the geodesic, we have the relation 
\begin{eqnarray}
&&X_0= R\sin t  = \frac{R}{M} \sqrt{\rho^2 -M^2} \sinh M \tau, \\ \nonumber
&&X_1 \cosh  \eta_1  + X_3\sinh \eta_1 =  R\cos t \cosh \eta_1  = \frac{R}{M} \rho \cosh M x |_{x=0}  ,  \\ \nonumber
&&X_2  =0  = \frac{R}{M} \rho  \sinh M x |_{x = 0}  \\ \nonumber
&& X_3 \cos\eta_1 + X_1 \sinh \eta_1 =  + R\cos t \sinh \eta_1  = \frac{R}{M} \sqrt{\rho^2 - M^2} \cosh M \tau 
\end{eqnarray}
From this, we see that  evaluating the charges for the geodesics in the BTZ embedding  can be done by relating the $SL(2, R)$ element along the geodesic in BTZ to that of $AdS_3$ as follows
as  on the geodesic
\begin{eqnarray} \label{groupelementr1}
&&g_{BTZ} =  
\left( 
\begin{array}{cc}
\cosh \frac{\eta_1}{2}   &  \sinh \frac{\eta_1}{2}  \\
\sinh\frac{\eta_1}{2} & \cosh \frac{\eta_1}{2} 
\end{array}
\right) 
g_{AdS_3} 
\left( 
\begin{array}{cc}
\cosh \frac{\eta_1}{2}   &  \sinh \frac{\eta_1}{2}  \\
\sinh\frac{\eta_1}{2} & \cosh \frac{\eta_1}{2} 
\end{array}
\right), 
\end{eqnarray}
where 
\begin{eqnarray}
&&g_{BTZ} = 
\left( 
\begin{array}{cc}
 \frac{2}{M} \rho \exp ( M x)  & 
\frac{2}{M} \sqrt{\rho^2 -M^2} \exp( M \tau)   \\
\frac{2}{M} \sqrt{\rho^2 -M^2} \exp( - M \tau)  &
\frac{2}{M}  \rho \exp (- M x)  
\end{array}
\right) |_{x=0}
\\ \nonumber
&& g_{AdS_3} 
= 
\left( 
\begin{array}{cc}
R \cos t  & R \sin t \\
-R \sin t &  R\cos t 
\end{array}
\right).
\end{eqnarray}
The sigma model in (\ref{sl2rsigma}) admits independent left and right $SL(2, R)$ symmetries. In (\ref{groupelementr1}), we have chosen to act by the same $SL(2, R)$ element both from the left and right. The left and right action on the group element $g_{AdS_3}$ are the same. From this relation, it is easy to see that evaluating the charges we obtain 
\begin{eqnarray}
Q_L^{(1)}|_{{\tiny BTZ}}   = 0,  \qquad 
Q_L^{(2)} |_{\tiny BTZ} =-\kappa \sinh\eta_1, \qquad Q_L^{(3)}|_{\tiny BTZ} =  \kappa \cosh\eta_1.
\end{eqnarray}

Proceeding similarly, we can evaluate the right $SL(2, R)$ charges, they are given by 
\begin{eqnarray} \label{defrighchg}
Q_R^{(1)}& =&   - \frac{  \dot \rho   }{\sqrt{ \rho^2 -M^2} }  \cosh M ( \tau +x)  + \frac{ ( \dot x - \dot \tau) }{M}
\rho \sqrt{ \rho^2 -M^2}  \sinh M ( \tau + x),  \\ \nonumber 
Q_R^{(2)} &=& \frac{  \rho^2 -M^2 }{M}  \dot \tau - \frac{\rho^2}{M} \dot x , \\ \nonumber
Q_R^{(3)}  &=&    \frac{  \dot \rho   }{\sqrt{ \rho^2 -M^2} }  \sinh M ( \tau + x)  + \frac{ ( \dot \tau - \dot x) }{M}
\rho \sqrt{ \rho^2 -M^2}  \cosh M ( \tau + x).
\end{eqnarray}
Again, substituting the initial conditions (\ref{initialcond})  in these charges, we obtain 
\begin{eqnarray} \label{rightcharge}
Q_R^{(1)}|_{{\tiny BTZ}}   = 0,  \qquad 
Q_R^{(2)} |_{\tiny BTZ} =\kappa \sinh\eta_1, \qquad Q_R^{(3)}|_{\tiny BTZ} =  \kappa \cosh\eta_1.
\end{eqnarray}
This of course is related to charge in $AdS_3$ by a $SO(1, 2)$ boost 
\begin{eqnarray} \label{boost2}
\left( \begin{array}{c}
Q_R^{(1)} \\ Q_R^{(2)} \\ Q_R^{(3)} 
\end{array}
\right)_{BTZ} = 
\left( \begin{array}{ccc}
1 & 0  & 0 \\
0 & \cosh\eta_1 & \sinh\eta_1 \\
0 & \sin\eta_1 & \cosh\eta_1 
\end{array}
\right) 
\left( \begin{array}{c}
Q_R^{(1)} \\ Q_R^{(2)} \\ Q_R^{(3)} 
\end{array}
\right)_{AdS_3}.
\end{eqnarray}
We can verify the result for the charges in  (\ref{rightcharge}) by using the relationship between the group elements of the geodesics in $AdS_3$ and BTZ  in (\ref{groupelementr}).

It is useful to evaluate the quadratic Casimir of these charges, which are given by 
\begin{eqnarray}\label{qcasimir}
Q_L\cdot Q_L = (Q_L^{(1)})^2 + (Q_L^{(2)})^2 - (Q_L^{(3)})^2 .
\end{eqnarray}
Evaluating this for the time-like geodesic in $AdS_3$, we obtain 
\begin{eqnarray}
Q_L\cdot Q_L|_{\tiny AdS_3} = - \kappa^2.
\end{eqnarray}
For the geodesic on BTZ, we obtain 
\begin{eqnarray} \label{casimir1}
Q_L\cdot Q_L|_{\tiny BTZ} 
&=&  \rho^2 \dot x^2  +\frac{\dot \rho^2 }{\rho^2 - M^2} -  (\rho^2 - M^2) \dot \tau^2 , \\ \nonumber
&=&  -\kappa^2.
\end{eqnarray}
To obtain the last line, we have substituted the initial conditions of the geodesic in (\ref{initialcond}). Similarly, the quadratic Casimir of the right charges is given by 
\begin{eqnarray}
Q_R\cdot Q_R = (Q_R^{(1)})^2 + (Q_R^{(2)})^2 - (Q_R^{(3)})^2 .
\end{eqnarray}
Evaluating this on the $AdS_3$ geodesic results in 
\begin{eqnarray}
Q_R\cdot Q_R|_{AdS_3} = - \kappa^2 .
\end{eqnarray}
For BTZ, we obtain 
\begin{eqnarray} \label{casimir2}
Q_R\cdot Q_R|_{\tiny BTZ} 
&=& \rho^2 \dot x^2  +\frac{\dot \rho^2 }{\rho^2 - M^2} -  (\rho^2 - M^2) \dot \tau^2 , \\ \nonumber
&=&  -\kappa^2.
\end{eqnarray}
As the charges for the  $AdS_3$ geodesic and BTZ are related by an $SO(1, 2)$ boost, the resulting Casimir is the same. This fact is useful to construct classical string solutions, as we will see subsequently. 

\subsection{In-falling particle with  longitudinal velocity} \label{sectionvel}

We now generalise the co-ordinate transformation in (\ref{coordinate transformation for in-falling geodesic}), the following
\begin{align}\label{coordinate transformation for boosted in-falling geodesic}
        &X_0 =\sqrt{r^2+R^2}\ \sin{t} =\frac{R}{M}\lrt{ \sqrt{\rho^2-M^2} \sinh{(M\tau)} \cosh{\eta_2} -\rho \sinh{\eta_2}\sinh{(Mx)} },  \nonumber\\
        &X_1=\sqrt{r^2+R^2} \cos{t} =\frac{R}{M}\lrt{ \rho \cosh{\eta_1} \cosh{(M x)} -\sqrt{\rho^2-M^2} \sinh\eta_1 \cosh{(M\tau)} },  \nonumber\\
        &X_2=r\sin{\phi}=\frac{R}{M} \lrt{\rho \sinh{(Mx)}\cosh{\eta_2} - \sqrt{\rho^2-M^2}\sinh{(M\tau)} \sinh{\eta_2}} , \\
    &X_3= r \cos{\phi}= \frac{R}{M}\lrt{\sqrt{\rho^2-M^2} \cosh{\eta_1} \cosh{(M\tau)} - \rho \sinh{\eta_1} \cosh{(Mx)}} .\nonumber
\end{align}
In comparison with the earlier transformation, we now have a boost  $\eta_2$ also in the  $X_0, X_2$ plane, along with the boost$\eta_1$  in the  $X_1, X_3$ plane. Let us study what happens to the time-like geodesic at the origin in the $AdS_3$ coordinates given by 
(\ref{origeod}). In the BTZ coordinates, this trajectory satisfies the condition
\begin{eqnarray} \label{vgeo0}
\rho \sinh( M x) \cosh \eta_2 &=& \sqrt{ \rho^2 -M^2} \sinh ( M \tau )  \sinh \eta_2, \\ \nonumber
\sqrt{\rho^2 -M^2} \cosh \eta_1 \cosh( M \tau) &=& \rho \sinh \eta_1 \cosh (M x) .
\end{eqnarray}
These relations arise from setting  $r=0$ in the last 2 equations of (\ref{coordinate transformation for boosted in-falling geodesic}). 
Now using these relations in the first equation of (\ref{coordinate transformation for boosted in-falling geodesic}), we obtain
\begin{eqnarray} \label{vgeo1}
R \sin t &=&\frac{R}{M} \sqrt{\rho^2 -M^2} \frac{ \sinh M \tau }{\cosh \eta_2} , \\ \nonumber
 &=&\frac{R}{M} \frac{\rho \sinh M x}{ \sinh \eta_2}, \\ \nonumber
 \end{eqnarray}
 From the  the 2nd equation of (\ref{coordinate transformation for boosted in-falling geodesic}), we obtain 
 \begin{eqnarray}\label{vgeo2}
 R\cos t &=& \frac{R}{M} \frac{\rho \cosh M x}{ \cosh \eta_1}, \\ \nonumber
 &=& \frac{R}{M}\sqrt{\rho^2 -M^2} \frac{\cosh M \tau}{\sinh \eta_1} .
\end{eqnarray}
From these equations we see that 
\begin{eqnarray} \label{circle}
\rho^2 
\left[  \Big(\frac{ \sinh M x }{\sinh \eta_2 } \Big)^2  + \Big( \frac{\cosh M x}{ \cosh \eta_1}  \Big)^2  \right] = M^2 , 
\\ \nonumber
(\rho^2  -M^2)
\left[  \Big(\frac{ \sinh M \tau }{\cos \eta_2 } \Big)^2  + \Big( \frac{\cosh M \tau}{ \sinh \eta_1}  \Big)^2  \right]
= M^2 .
\end{eqnarray}

\subsubsection*{Constants of motion along geodesics}

The equations (\ref{vgeo0}), (\ref{vgeo1}), (\ref{vgeo2}) describe the trajectory of an in-falling particle in BTZ but with velocity in the $x$ direction. 
To see this, we differentiate the 2nd equation of (\ref{vgeo1}) with respect to $\sigma^0$  and use 
(\ref{vgeo2}) to obtain 
\begin{eqnarray} \label{dgeo1}
\frac{1}{\sinh \eta_2} \Big( \dot \rho \sinh M x   + \rho M \dot x \cosh M x \Big) = \kappa \rho \frac{\cosh M x}{\cosh \eta_1} . 
\end{eqnarray}
Similarly differentiating the first equation of (\ref{vgeo2}) and using the second equation of (\ref{vgeo1})
we get
\begin{eqnarray} \label{dgeo2}
\frac{1}{\cosh \eta_1} \Big( \dot \rho \cosh M x + \rho M \dot x \sinh M x \Big)  = - \kappa \rho \frac{\sinh M x}{ \sinh \eta_2}.
\end{eqnarray}
Solving for $\dot x$ from (\ref{dgeo1}) and (\ref{dgeo2}) and using (\ref{circle}), we obtain 
\begin{eqnarray} \label{c1}
 \rho^2 \dot x = M \kappa \cosh \eta_1 \sinh \eta_2.
\end{eqnarray}
Comparing with (\ref{constants}), we see that this is clearly the equation for the constant of motion for the momentum along the $x$-direction for a geodesic 
in the BTZ geometry. Note that the velocity $\dot x$ is non-zero only when the boost parameter $\eta_2$ does not vanish. Proceeding along similar lines, we can differentiate the first equation in (\ref{vgeo1}) and the second equation in (\ref{vgeo2}) to obtain 
\begin{eqnarray} \nonumber
\frac{1}{\cosh \eta_2} \Big( 
\frac{\rho\dot\rho }{\sqrt{ \rho^2 - M^2} } \sinh M\tau +  M \sqrt{\rho^2 - M^2}  \dot \tau \cosh M \tau \Big) 
= \kappa \sqrt{\rho^2 - M^2} \frac{\cosh M \tau }{ \sinh \eta_1} , \\  \nonumber
\frac{1}{\sinh \eta_1} \Big( 
\frac{\rho\dot\rho }{\sqrt{ \rho^2 - M^2} } \cosh M\tau + M \sqrt{\rho^2 - M^2} \dot \tau \sinh M \tau \Big) 
= -\kappa \sqrt{\rho^2 - M^2} \frac{\sinh  M \tau }{ \cosh \eta_2}
\\  
\end{eqnarray}
Solving for $\dot \tau$ from  these equations and using the second equation in (\ref{circle}), we obtain 
\begin{eqnarray} \label{c2}
( \rho^2 - M^2) \dot \tau = M \kappa \cosh \eta_2 \sinh\eta_1.
\end{eqnarray}
This is the constant of motion due to the time translation symmetry of the BTZ metric (\ref{constants}). 
Now taking the ratio of (\ref{c1}), (\ref{c2}) we obtain 
\begin{eqnarray} \label{ratiovel}
\frac{\rho^2}{\rho^2 -M^2} \dot x  =\tanh \eta_2 \coth \eta_1 \dot \tau .
\end{eqnarray}

\subsubsection*{Radial equation along the geodesic}

We now need to find the geodesic equation for the radial coordinate. 
This involves several manipulations with the basic equations relating the trajectory in $AdS_3$ and BTZ. 
From  the 2 equations in (\ref{vgeo0}), we can solve for $\cosh M x$, 
\begin{eqnarray} \label{solcosh}
(\cosh M x)^2 = \frac{1}{\rho^2} \frac{ ( \rho^2 - M^2) ( \tanh \eta_2)^2 - \rho^2 }{ (\tanh \eta_2 \tanh \eta_1)^2 - 1} .
\end{eqnarray}
Using this in the second equation of (\ref{vgeo0}), we also have 
\begin{eqnarray} \label{solct}
(\cosh M \tau)^2 = \frac{(\tanh \eta_1)^2}{ \rho^2 - M^2 } \frac{ ( \rho^2 - M^2) ( \tanh \eta_2 )^2- \rho^2 }{
( \tanh \eta_2 \tanh \eta_1)^2  - 1} .
\end{eqnarray}
Differentiating the second equation in (\ref{vgeo0}) and eliminating  the $\dot x $ term using (\ref{ratiovel}), we obtain 
\begin{eqnarray}
\sinh M \tau   \Big( 1- \frac{\rho^2 - M^2}{\rho^2} (\tanh \eta_1)^2  \Big)  \dot \tau
= - \frac{M \dot \rho}{ ( \rho^2 - M^2)^{\frac{3}{2} } }\tanh \eta_1 \cosh M x.
\end{eqnarray}
By squaring this equation and using (\ref{solcosh}), we can write 
\begin{eqnarray}
(\sinh M \tau)^2   \dot \tau^2 &=&  - \frac{M^2 \dot \rho^2}{ ( \rho^2 - M^2) }   \\ \nonumber
&&  \times 
\frac{\rho^2}{ \rho^2 - ( \rho^2 - M^2) \tanh^2 \eta_2 } \times  \frac{\tanh^2 \eta_1}{\tanh^2 \eta_2 \tanh^2 \eta_1 - 1} .
\end{eqnarray}
We can eliminate $\dot \tau$ using the  conservation equation (\ref{c2}), to obtain 
\begin{eqnarray} \label{solst}
\kappa^2 \sinh^2 M \tau &=& - \frac{\dot \rho ^2}{ \rho^2 - M^2 }  \\ \nonumber
&& \times \frac{1}{ \sinh^2 \eta_2 \sinh^2 \eta_1 - \cosh^2 \eta_2 \cosh^2 \eta_1}  \times
\frac{\rho^2}{ \rho^2 - ( \rho^2 - M^2) \tanh^2 M \eta_2 } .
\end{eqnarray}
Now using the (\ref{solct}) and (\ref{solst}), we can eliminate the $\tau$ dependence to obtain
\begin{eqnarray}
&& -\frac{ \kappa^2}{\rho^2}  \Big[ \rho^2 - ( \rho^2 -M^2) ( \tanh \eta_2)^2  \Big]
\Big[ ( \sinh  \eta_2 \sinh \eta_1)^2  -  (\cosh \eta_2 \cosh \eta_1)^2  \Big]  \\ \nonumber
&& \qquad\qquad= \frac{ \kappa^2}{ \rho^2 ( \rho^2 - M^2) } 
( \sinh \eta_1 \cosh \eta_2 )^2 
\Big[  ( \rho^2 - M^2 )( \tanh\eta_2)^2  - \rho^2 \Big]^2 
- \frac{\dot \rho^2}{ \rho^2 -M^2}.
\end{eqnarray}
Opening up the terms and re-grouping, we obtain 
\begin{eqnarray} \label{radialeq}
-\frac{\dot \rho^2}{ \rho^2 -M^2} -\frac{ ( \kappa M\sinh \eta_2 \cosh \eta_1)^2 }{\rho^2}
+ \frac{( \kappa M  \sinh \eta_1 \cosh  \eta_2)^2 }{ \rho^2 -M^2} 
= \kappa^2 .
\end{eqnarray}
We see that this  equation coincides the radial equation of the geodesic in BTZ given in (\ref{constants3}). Using (\ref{c1}), ( \ref{c2}) and (\ref{radialeq}), we can identify the constants of motion to be 
 \begin{eqnarray}
 \frac{E}{R^2} = -\kappa M \sinh\eta_1 \cosh\eta_2,  \qquad 
  \frac{ v}{R^2} =  M \kappa \cosh\eta_1 \sinh\eta_2,  \qquad
  \frac{\tilde m^2}{R^2} = \kappa^2 .
 \end{eqnarray}

\subsubsection*{Left and right $SL(2, R) $ charges}

We can also find the initial conditions at $\sigma^0=0$, when the geodesic has zero radial velocity
 $\dot \rho( 0) = 0$. 
\begin{eqnarray}
& & x(0) = 0, \quad \dot x (0 ) =  \frac{\kappa}{M} \frac{\sinh \eta_2}{\cosh \eta_1}, \qquad\qquad
\tau(0) = 0, \quad \dot \tau(0) = \frac{\kappa}{M} \frac{\cosh\eta_2}{\sinh\eta_1}.  \\ \nonumber
&& \rho(0) = M \cosh\eta_1, \quad \dot \rho(0) = 0 .
\end{eqnarray}
These initial conditions are obtained by solving (\ref{radialeq}) and (\ref{circle}), together with the 
conservation equations (\ref{c1}), (\ref{c2}). 
With these initial conditions, we can find the left and right $SL(2, R)$ charges in (\ref{defleftchg}), (\ref{defrighchg}). 
For the left charges, we obtain 
\begin{eqnarray} \label{leftvcharge}
Q_L^{(1)}|_{BTZ} = 0, \qquad Q_L^{(2)}|_{BTZ} = - \kappa \sinh(\eta_1 + \eta_2), \qquad  Q_R^{(1)}|_{BTZ}= \kappa \cosh(\eta_1+\eta_2) . \nonumber
\\
\end{eqnarray}
Similarly, evaluating the right charges, we get
\begin{eqnarray} \label{rightvcharge}
Q_R^{(1)}|_{BTZ} = 0, \qquad Q_R^{(2)}|_{BTZ} =  \kappa \sinh(\eta_1 - \eta_2), \qquad  Q_R^{(1)}|_{BTZ}= \kappa \cosh(\eta_1-\eta_2) . \nonumber
\\
\end{eqnarray}
It is clear that the $Q_L|_{BTZ}, Q_R|_{BTZ}$  
charges for the geodesic with velocity can be obtained from the charges of the time-like geodesic at the origin of $AdS_3$ by the boosts in (\ref{boost1}),(\ref{boost2}) by replacing the boost parameter to $\eta_1\rightarrow \eta_1+\eta_2$ and  $\eta_1 \rightarrow \eta_1-\eta_2$ respectively. This fact can also be seen, from the transformation in (\ref{coordinate transformation for boosted in-falling geodesic}), on the geodesic we have the relations
\begin{eqnarray}
&&X_0\cosh \eta_2 + X_2 \sinh \eta_2= R\sin t  \cos \eta_2  = \frac{R}{M} \sqrt{\rho^2 -M^2} \sinh M \tau, \\ \nonumber
&&X_1 \cosh  \eta_1  + X_3\sinh \eta_1 =  R\cos t \cosh \eta_1  = \frac{R}{M} \rho \cosh M x   ,  \\ \nonumber
&&X_2  \cosh\eta_2 + X_0\sinh\eta_2 =R \sin t \sinh\eta_2 = \frac{R}{M} \rho  \sinh M x ,  \\ \nonumber
&& X_3 \cos\eta_1 + X_1 \sinh \eta_1 =   R\cos t \sinh \eta_1  = \frac{R}{M} \sqrt{\rho^2 - M^2} \cosh M \tau .
\end{eqnarray}
On the geodesic, the $SL(2, R)$  element  parameterised by BTZ co-ordinates is related to that of $AdS_3$ by 
\begin{eqnarray} \label{groupelementr}
&&g_{BTZ} =  
\left( 
\begin{array}{cc}
\cosh \frac{\eta_1 +\eta_2 }{2}   &  \sinh \frac{\eta_1 +\eta_2}{2}  \\
\sinh\frac{\eta_1 +\eta_2 }{2} & \cosh \frac{\eta_1 +\eta_2}{2} 
\end{array}
\right) 
g_{AdS_3} 
\left( 
\begin{array}{cc}
\cosh \frac{\eta_1-\eta_2}{2}   &  \sinh \frac{\eta_1-\eta_2}{2}  \\
\sinh\frac{\eta_1-\eta_2}{2} & \cosh \frac{\eta_1-\eta_2}{2} 
\end{array}
\right), 
\end{eqnarray}
where 
\begin{eqnarray}
&&g_{BTZ} = 
\left( 
\begin{array}{cc}
 \frac{2}{M} \rho \exp ( M x)  & 
\frac{2}{M} \sqrt{\rho^2 -M^2} \exp( M \tau)   \\
\frac{2}{M} \sqrt{\rho^2 -M^2} \exp( - M \tau)  &
\frac{2}{M}  \rho \exp (- M x)  
\end{array}
\right), 
\\ \nonumber
&& g_{AdS_3} 
= 
\left( 
\begin{array}{cc}
R \cos t  & R \sin t \\
-R \sin t &  R\cos t 
\end{array}
\right).
\end{eqnarray}
This relation between the group elements also enables the easy evaluation of the charges, and we obtain the results (\ref{leftvcharge}), (\ref{rightvcharge}) for the left and right $SL(2, R)$  charges. Since the left and the right $SL(2, R)$ charges are related by a boost, the quadratic Casimir of these charges remains invariant, and the relations (\ref{casimir1}), (\ref{casimir2}) continue to hold.

\section{In-falling classical strings in BTZ} \label{sec3}

In $AdS_3 \times S^3 \times M $ several classical string solutions are constructed as follows:
One considers a time like geodesic at the origin of $AdS_3$, and a classical solution of the sigma model on  $S^3$ such that the solution satisfies the classical Virasoro condition. To proceed further,  we describe some aspects of the $SU(2)$ WZW  model, it is given by the action 
\begin{eqnarray} \nonumber
  S_{SU(2)}  &=&-\frac{k}{8\pi}\int_{\Sigma} d^2\sigma\ Tr\lrt{\hat g^{-1}\partial_a \hat g\ \hat g^{-1}\partial^a \hat g} 
   + \frac{ k}{12\pi}\ q \int_{{\cal B}} Tr\lrt{\hat g^{-1}d \hat g \wedge \hat g^{-1}d\hat g \wedge \hat g^{-1}d\hat g} .
  \\ 
\end{eqnarray}
Here $\hat g \in SU(2)$ which can be expressed as 
\begin{align}
g= \begin{pmatrix}
      Z_1 & Z_2\\
      -Z_2^* & Z_1^*
    \end{pmatrix}, 
    \quad \quad |Z_1|^2 + |Z_2|^2= 1. 
\end{align}
Again $q$ lies between $0$ and $1$. It is $0$ for the background with pure RR flux and $1$ for the pure NS flux and $\alpha' k = R^2$ \footnote{In case $M = S^3 \times S^1$, then the levels of the 2  $SU(2)$ WZW models are related to the level of the $SL(2, R)$ WZW model. For eg. if $q= 1$, then $\frac{1}{k} = \frac{1}{k_1} + \frac{1}{k_2}$ where $k_1, k_2$ are the levels of the 2 $SU(2)$ models \cite{Elitzur:1998mm}. }.
Classically, the model admits $SU(2)_L,  SU(2)R$ conserved charges which are given by 
\begin{eqnarray} \label{defcharge}
\hat Q_L = \frac{k}{2\pi } \int_0^{2\pi} d\sigma^1 \Big( \hat g^{-1}\dot{\hat g} + q \,  \hat g^{-1} \hat g'  \Big) ,
\qquad 
\hat Q_R =  \frac{k}{2\pi } \int_0^{2\pi } d\sigma^1 \Big(   \dot{\hat g}  \hat g^{-1}  - q \,  \hat g' \hat g^{-1} \Big) .
\end{eqnarray}
We can project these charges along the $3$ generators of $SU(2)$ by defining
\begin{align}\label{defcharge2}
    J_L^i= i   \ {\rm Tr} [\mathcal{\varsigma}^i.  \hat Q_L], \quad J_R^i= i \ {\rm Tr} [\mathcal{\varsigma}^i.  \hat Q_R],
\end{align}
where 
\begin{eqnarray}
\mathcal{\varsigma}^1= \frac{1}{2} \sigma^1,\qquad \mathcal{\varsigma}^2= \frac{1}{2}\sigma^2, \qquad \mathcal{\varsigma}^3= \frac{1}{2} \sigma^3.
\end{eqnarray}
To make matters more explicit, it is useful to have a parametrisation of the group element in 
$SU(2)$ by 
\begin{align}\label{parameterization of su2 group element}
    Z_1 = X_1 + i X_2= \sin\psi\ e^{i\phi_1}, \quad Z_2 = X_3 + i X_4= \cos\psi\  e^{i\phi_2}.
\end{align}
The metric on the sphere, then, is given by 
\begin{eqnarray} \label{mets3}
ds^2_{S^3} = d\psi^2 + \sin^2\psi d\phi_1^2 + \cos^2 \psi d\phi_2^2.
\end{eqnarray}
Let us choose the $B_{NS}$ field to be given by \footnote{ An additive constant to the $B_{NS}$ field 
shifts the sigma model action by a total derivative.
We have chosen this constant $c$ defined in \cite{Hoare:2013lja} to be $c=-1$. } 
  \begin{eqnarray}
  B_{\phi_1\phi_2} = q \sin^2 \psi = \frac{q }{2} ( \cos 2 \psi  - 1) .
  \end{eqnarray}
 We can add a constant to the field $B$, and it does not affect the equations of motion. 
Quite often, it is convenient to choose  the following combination of angular momenta 
$J_1$ and $J_2$ which generate the shift in $\phi_1$ and $\phi_2$
\begin{align}
  J_1= -\frac{1}{2}\lr{ J^3_R + J^3_L },\quad  J_2 =   -\frac{1}{2}\lr{  J^3_R- J^3_L }.
\end{align}

Consider any classical solution in $S^3$, say, geodesics on $S^3$,  circular strings and giant magnons moving on a time-like geodesic at the origin on $AdS_3$ and at a fixed point on $M$. These are solutions to the equation of motion of the sigma model; they are valid classical solutions of the string theory as long as the Virasoro constraints are satisfied. 
 \begin{eqnarray}\label{vircons}
&&T_{00} ( AdS_3)   + T_{00} (S^3)  + T_{00} (M)= 0,  \\ \nonumber
&&T_{01} ( AdS_3)   + T_{01} (S^3)  + T_{01} (M)= 0.
\end{eqnarray}
Here  we have used $T_{00} = T_{11}, 
T_{01} = T_{10}$ to write down the non-trivial constraints. 
Geodesics, circular strings and giant magnons satisfy the constraint
$T_{01} ( S^3)  =0$, 
when evaluated on the respective classical solutions, as we will review subsequently. 
There is no contribution to the world sheet stress tensor from the sigma model on $M$ since the string is 
at a fixed position on $M$, therefore we have  $T_{01} ( M)  =0$. 
Finally, since the solution is the time-like geodesic  at the origin of $AdS_3$ with no dependence on the world sheet spatial direction, we  have 
$T_{01} ( AdS_3)  =0$. 
This implies the 2nd Virasoro constraint in (\ref{vircons}) is satisfied. 
Now, the first Virasoro constraint on these solutions reduces to 
\begin{eqnarray}
-k \kappa^2 + T_{00}( S^3) = 0.
\end{eqnarray}

This discussion and the conclusions of section \ref{sec2} then imply the following. Once we map the time-like geodesic to the in-falling geodesic in BTZ, it is guaranteed that the sigma model equations of motion in $BTZ \times S^3 \times M$ are satisfied. We can also see that the following Virasoro constraints are satisfied
\begin{eqnarray}\label{vircons2}
T_{00} ( BTZ)   + T_{00} (S^3)  + T_{00} (M)= 0,  \\ \nonumber
T_{01} ( BTZ)   + T_{01} (S^3)  + T_{01} (M)= 0.
\end{eqnarray}
The 2nd Virasoro constraint in (\ref{vircons2}) is satisfied, since we still do not have world sheet spatial dependence in the $BTZ$ and the world sheet momentum of each of these solutions vanish on $S^3$ and $M$. Finally,  the first Virasoro constraint is satisfied due to the non-trivial observation in the previous section.
The maps (\ref{coordinate transformation for in-falling geodesic}), (\ref{coordinate transformation for boosted in-falling geodesic}), which take the time-like geodesic at the origin of $AdS_3$ to the in-falling geodesic in BTZ is such that the $SL(2, R)$ Casimir $\kappa^2$ is invariant as shown for example in  (\ref{casimir1}), (\ref{casimir2}). The world sheet stress tensor in BTZ on solutions, which depends only on world sheet time, is  given by 
\begin{eqnarray} \label{btzstressca}
T_{00}(BTZ) &=& k \left( - \frac{E^2}{ R^4 ( \rho^2 - M^2) } + \frac{\dot \rho^2}{ \rho^2 - M^2} 
+ \frac{v^2}{ R^4 \rho^2}  \right) , \\ \nonumber
&=&  kQ_L\cdot Q_L =  kQ_R\cdot Q_R,  \\ \nonumber
&=& - k \kappa^2.
\end{eqnarray}
To obtain these results, we have used the identification of the quadratic Casimir with the world sheet stress tensor in BTZ from (\ref{casimir1}), (\ref{casimir2}). 
This then implies that the first Virasoro constraint (\ref{vircons2})  is satisfied if the solution satisfies the
corresponding solution on $AdS_3$. Therefore, we conclude that the map which takes the time like geodesic at the origin of $AdS_3$ also takes
classical string solutions from $AdS_3\times S^3 \times M$ to $BTZ\times S^3 \times M$. As we have emphasised, these classical string solutions involve geodesics in $AdS_3$ and BTZ. To make matters concrete, we will discuss the case of geodesics, circular strings and giant magnons on $S^3$ falling in BTZ.

\subsubsection*{Geodesics}
  
  Geodesics on $S^3$ falling into the BTZ horizon are given by 
  \begin{eqnarray} \label{fullgeo}
  && \rho(\sigma^0, \sigma^1 ) = \rho( \sigma^0) ,
   \qquad \tau (\sigma^0, \sigma^1 )= \tau( \sigma^0) , \qquad x (\sigma^0, \sigma^1 )  = x(\sigma^0),  \\ \nonumber
  && \psi (\sigma^0, \sigma^1 ) =  0 ,  \qquad \phi_1 ( \sigma^0, \sigma^1)  =  0,
  \qquad \phi_2 ( \sigma^0, \sigma^1)  =  \frac{J}{k} \sigma^0 , \\ \nonumber
  && X(\sigma^0, \sigma^1)|_M = {\rm constant} .
  \end{eqnarray}
  where the BTZ co-ordinates $\{\rho, \tau, \sigma \}$ 
    depend  only on the  world sheet time and are obtained by the maps 
    (\ref{coordinate transformation for in-falling geodesic}), 
    (\ref{coordinate transformation for boosted in-falling geodesic})
   and therefore, they satisfy the general geodesic equations (\ref{c1}), (\ref{c2}), (\ref{radialeq}). 
   Let us evaluate the $SU(2)$ charges using (\ref{defcharge2}). 
   \begin{eqnarray}
    J_{L,R}^1 =0, \quad J_{L,R}^2 =0,\quad J_{L}^3= J,\quad J^3_{R}=-J.
   \end{eqnarray}
   It is clear that the solution in (\ref{fullgeo}) satisfies the 2nd Virasoro constraint  (\ref{vircons2}). 
   On evaluating the first Virasoro constraint, we obtain 
   \begin{eqnarray}
   -k \kappa^2  + \frac{1}{k} (J)^2 = 0 , \qquad T_{00}( S^3) = \frac{(J)^2}{k} .
\end{eqnarray}
We identify  $\Delta = k \kappa $ and therefore, we obtain the BPS relation
\begin{eqnarray}
\Delta = J.
\end{eqnarray}
It is important to realise that these solutions are geodesics falling into the horizon of the BTZ. 
  
  \subsubsection*{Circular Strings}

  The in-falling circular string in BTZ is given by 
  \begin{eqnarray} \nonumber
    && \rho(\sigma^0, \sigma^1 ) = \rho( \sigma^0) ,
   \qquad \tau (\sigma^0, \sigma^1 )= \tau( \sigma^0) , \qquad x (\sigma^0, \sigma^1 )  = x(\sigma^0),  \\ \nonumber
  &&  Z_1= \sqrt{\frac{W- q m}{2W}}\ e^{i\lrt{\sigma^0\lr{W + qm} +m\sigma^1 }}\ ,\quad\qquad Z_2= \sqrt{\frac{W+q m}{2W}}\ e^{i\lrt{\sigma^0\lr{W - qm} -m\sigma^1}},\\ \nonumber
   && W=\sqrt{w^2 + q^2m^2}, \\
   && X(\sigma^0, \sigma^1)|_M = {\rm constant} .
\end{eqnarray}
Here $m \in \mathbb{Z}$, the trajectory in BTZ is that of a geodesic given by the equations  (\ref{c1}), (\ref{c2}), (\ref{radialeq}), while the solution on the $S^3$ is the circular string solution found by  \cite{Hoare:2013lja}  for the mixed Ramond-Ramond and NS flux, $q\neq 0$. The first Virasoro constraint  in (\ref{vircons})  is satisfied provided
\begin{eqnarray}
\kappa^2 = \omega^2 + m^2 .
\end{eqnarray}
We have again used equation (\ref{btzstressca}) to evaluate the stress tensor of the solution in the BTZ direction. It can be seen that the second Virasoro is also satisfied. Evaluating the angular momenta of the solution  on $S^3$, we obtain 
\begin{eqnarray}
    J_1 = J_2= \frac{k}{2}\lr{W - qm}.
\end{eqnarray}
To complete the discussion, we write down the dispersion relations
\begin{eqnarray}
\Delta = k \kappa =  \sqrt{( J + qkm)^2 + k^2m^2(1-q^2)}, \quad \quad J\equiv J_1 + J_2= 2J_1.
\end{eqnarray}
Expanding in the large $J$ limit:
\begin{align}
    \Delta - J=  qkm + \frac{k^2 m^2(1-q^2)}{2J} + \mathcal{O}(J^{-2}).
\end{align}
Again, we mention that the maps (\ref{coordinate transformation for in-falling geodesic}),(\ref{coordinate transformation for boosted in-falling geodesic}), which take the time-like geodesic at the origin in $AdS_3$ to the in-falling geodesic in BTZ, ensure that the circular string solution in $AdS_3$  is a solution in BTZ with the same dispersion relation. We need to think of $\Delta = k \kappa$ as the quadratic Casimir in the $SL(2, R)$  WZW model.

\subsection*{Giant Magnon}

For these classes of solutions, we consider the case of $q=0$ and the simplest giant magnon solution.  The discussion can be easily generalised for more 
general giant magnons found in  \cite{Chen_2006} and for the solutions with $q\neq 0$ in \cite{Hoare:2013lja} as we will outline subsequently. The simplest giant magnon solution in BTZ  is obtained by the following string configuration
\begin{eqnarray} \label{magnsoln}
  && \rho(\sigma^0, \sigma^1 ) = \rho( \sigma^0), 
   \qquad \tau (\sigma^0, \sigma^1 )= \tau( \sigma^0) , \qquad x (\sigma^0, \sigma^1 )  = x(\sigma^0),  \\ \nonumber    &&Z_1= \lrt{\sin{\frac{p}{2}} \tanh{y} -i \cos{\frac{p}{2}} } e^{i\tilde{\sigma}^0},\quad Z_2=\frac{\sin{\frac{p}{2}}}{\cosh{y}}, \\ \nonumber
   && \quad y=\frac{\tilde{\sigma}^1- \tilde{\sigma}^0 \cos{\frac{p}{2}} }{\sin{\frac{p}{2}}},  
   \qquad \tilde{\sigma}^0 = \kappa \sigma^0, \quad  \tilde \sigma^1 = \kappa \sigma^1.
   \end{eqnarray}
   The $S^3$ part of the solution was found first in \cite{Hofman_2006} and then discussed in 
   detail in \cite{Chen_2006}. 
   It can be verified that the solution satisfies the first Virasoro condition in (\ref{vircons2}), again, the reason this happens is because of the fact that the geodesic in BTZ satisfies the equation (\ref{btzstressca}). 
   Let us evaluate the charges 
   \begin{eqnarray} \label{magdelta}
   \Delta &=& \frac{k}{2\pi } \int _{-\pi}^\pi  d\sigma^1 \sqrt{ - Q_L \cdot Q_L } , \\ \nonumber
   &=&  k \kappa ,
   \end{eqnarray}
   where $Q_L \cdot Q_L$ is the $SL(2, R)$ Casimir defined in (\ref{qcasimir}). 
   The angular momenta of this solution are given by 
   \begin{eqnarray} \label{magcharge}
       J_1 = \frac{k }{2\pi } \int_{-\pi}^\pi  d\sigma^1 {\rm Im}
       \big[ Z_1^*\, \partial_{\sigma^0} Z_1 \big], 
       \qquad J_2 = \frac{k}{2\pi } \int_{-\pi}^\pi  d\sigma^1 {\rm Im }  [ Z_2^*\, \partial_{\sigma^0}  Z_2].
   \end{eqnarray}
   The giant magnon solution is obtained by taking the scaling limit
   \begin{eqnarray} \label{maglim}
   \kappa,\Delta, J_1 \rightarrow\infty \quad {\rm with} \quad  k, J_2, (\Delta-J_1), \sigma_0, \sigma_1  \quad {\rm fixed. }
   \end{eqnarray}
   Evaluating the charges in this limit, we obtain
   \begin{eqnarray} \label{dispersionm}
   \Delta - J_1  &=& \frac{k}{2\pi} \int_{-\infty}^\infty d\tilde \sigma^1  
   \Big( 1- {\rm Im} \big[ Z_1^* \, \partial_{\tilde\sigma^1} Z_1\big]  \Big) , \\ \nonumber
   J_2 &= & 0 ,
   \end{eqnarray}
   where we have used  (\ref{btzstressca}). 
   \begin{eqnarray}
   \frac{\sqrt{- Q_L \cdot Q_L }}{\kappa } =1.
   \end{eqnarray}
   Substituting the solution (\ref{magnsoln}) in (\ref{dispersionm}), we obtain 
   \begin{eqnarray}
   \Delta - J_1 = \frac{k}{\pi} \sin \frac{p}{2}.
 \end{eqnarray}
 We emphasise that the dispersion relation is again identical to that of the magnon in $AdS_3\times S^3$, with 
 $\Delta$ being identified as the quadratic Casimir in the $SL(2 , R)$  WZW model. 
 
 For completeness, we mention that the boundary conditions that the solution in (\ref{magnsoln}) satisfies in the 
 limit (\ref{maglim})  is given by 
 \begin{eqnarray}
 \lim_{\tilde \sigma_1 \rightarrow \pm \infty}  Z_1 = e^{ i ( \tilde \sigma_0-\frac{\pi}{2}  \pm \frac{p}{2} ) } , 
 \qquad 
  \lim_{\tilde \sigma_1 \rightarrow \pm \infty} Z_2 = 0. 
 \end{eqnarray}
 Therefore, the ends of the strings are at $\psi = \frac{\pi}{2}$ but at angles such that $\delta  \phi_1 = p$.  Considering several giant magnons, whose momentum is such that their sum vanishes, the second Virasoro condition in (\ref{vircons2}) can be satisfied.  This is the same way the Virasoro condition for the case of magnons on $AdS_3 \times S^3$ is satisfied. It is clear that the  embedding of the magnon in BTZ given in (\ref{magnsoln}) can be extended  for 
 the general  dyonic magnon solutions with charge $J_2\neq 0$  found in \cite{Chen_2006} which obey the  dispersion relation 
 \begin{eqnarray}
 \Delta-J_1 =\sqrt{ J_2^2 + \frac{k^2}{\pi^2} \sin^2{\frac{p}{2}}}.
 \end{eqnarray}
 We can also generalise the embedding for the giant magnon solutions in the presence of 
 NS flux $q\neq 0$,  found in 
 \cite{Hoare:2013lja}.  All we have to do is take the BTZ part of the solution to be the in-falling geodesic 
 obeying the trajectory  (\ref{c1}), (\ref{c2}), (\ref{radialeq})
 and the $S^3$ part to be that constructed in the earlier works.

\section{Plane wave spectra in BTZ}

In section \ref{sec3}, we demonstrated that the classical solutions of the string sigma model 
$AdS_3\times S^3 \times M$ can be mapped to solutions in $BTZ \times S^3 \times M$. 
These solutions involve the timelike geodesics at the origin of  $AdS_3$, which are mapped to an in-falling geodesic in BTZ. We also showed that the solutions obey an identical dispersion relation once the conformal dimension $\Delta^2$ is identified with the quadratic Casimir of the WZW model on BTZ. The plane wave limit of $AdS_3\times S^3 \times M$  \cite{Berenstein:2002jq}, involves a time-like geodesic at the origin with angular momentum along one of the circles of $S^3$. The observations in section \ref{sec3}  suggest that it should be possible to obtain the same plane wave geometry in $BTZ\times S^3 \times M$ by zooming in on the geometry around in-falling geodesics in BTZ. Contrary to the above argument, it is known that Penrose limit around generic geodesics in black hole backgrounds leads to time-dependent singular plane waves \cite{Blau:2003dz, Blau:2004yi}. Let us briefly review these results to place the observation for the Penrose limit for in-falling geodesics in the BTZ black hole in a relevant context. Consider the Schwarzschild black hole in $AdS_5 \times S^5$, the metric is given in 
\begin{eqnarray}\label{AdS5S5BhMetric}
ds^2 = - \Big( r^2 + 1 - \frac{r_0^2}{r^2} \Big) dt^2 + \frac{dr^2}{ r^2 + 1 - \frac{r_0^2}{r^2} }+ r^2 ( d\phi^2 + \sin^2 \phi d\Omega_3^2) 
+ d\psi^2 + \sin^2 \psi d\Omega_4^3 .
\end{eqnarray}
Consider the case of a  null geodesic carrying angular momentum along $\phi$, 
 one of the circles in the $AdS_5$ directions. The trajectory of this geodesic is governed by  the following equations
 \begin{eqnarray} \label{schbhgeo}
 & &\Big( r^2 +1 - \frac{r_0^2}{r^2} \Big)  \dot t = E, \qquad\qquad r^2 \dot\phi = J , \\ \nonumber
 & &\dot r^2 + \frac{ J^2 }{r^2} \Big( r^2 + 1 - \frac{r_0^2}{r^2} \Big) = E^2 , \qquad 
 \psi = \frac{\pi}{2}.
 \end{eqnarray}
 The procedure for taking the Penrose limit along this geodesic has been detailed in Appendix \ref{A:penrose limit of AdS5S5}. This  results in the following pp wave metric in Brinkmann coordinates
 \begin{eqnarray}\label{planewavesch}
 & &ds^2 = 2 dx^+ dx^- +  \Big[ A x_1^2 + B ( x_2^2 + x_3^3 ) \Big] (dx^+)^2 + 
 \sum_{i =1}^3 dx_i^2 + \sum_{a = 1}^5  dy_a^2 , \\ \nonumber
 & &A = \frac{4 J^2 r_0^2}{r^6(x^+) }, \qquad B  = - \frac{2 J^2 r_0^2}{ r^6(x^+) }.
 \end{eqnarray}
 In the coefficients $A, B$, we need to substitute for the radial coordinate $r$ in terms of the affine parameter $x^+$ by solving the second equation in (\ref{schbhgeo}). 
 The coordinates $y_a$ originate from the sphere $S^5$.  Note that for a radially in-falling geodesic $J=0$, the plane wave geometry (\ref{planewavesch}) reduces to flat space. Thus, the geometry in general depends on time $x^+$ and has been shown to be singular at $x^+$ such that $r(x^+) =0$  \cite{Blau:2003dz, Blau:2004yi}. 
 Let us examine the situation in which the null geodesic has angular momentum along one of the angles of $S^5$. 
 The trajectory of this geodesic is determined by the equations 
 \begin{eqnarray} \label{geoschspher}
 \Big( r^2 + 1 - \frac{r_0^2}{r^2}  \Big) \dot t = E, \qquad \qquad  \dot \psi = J , \\ \nonumber
 \dot r^2 + J^2 \Big( r^2 +1 - \frac{r_0^2}{r^2} \Big) = E^2, \qquad \qquad \phi =\frac{\pi}{2}.
 \end{eqnarray}
 Zooming near this geodesic, the geometry reduces to the following plane wave metric
 \begin{eqnarray} \label{ppwavelimscha}
    ds^2 = 2 dx^+dx^- - J^2\lrt{\big( 1-\frac{3r_0^4}{r^4} \big)  x_1^2 + \big( 1 + \frac{r_0^4}{r^4} \big) \vec{z_3}^2 +
   \vec{z_4}^2}\lr{dx^+}^2 +  dx_1^2 + 
   d\vec{z_3}^2+ d\vec{z_4}^2 ,\nonumber \\
\end{eqnarray}
where $\vec{z_3}$ refers to the 3 co-ordinates that  label $\mathbb{R}^3$ and 
 $\vec{z_4}$ refers to the 4 co-ordinates that label $\mathbb{R}^4$. The details of obtaining the Penrose limit, together with the value of the $5$ form fluxes, are given in appendix \ref{A:penrose limit of AdS5S5}. Here again, the radial coordinate is a function of light cone time $x^+$, which is determined by solving the radial equation in (\ref{geoschspher}). This implies the metric  (\ref{ppwavelimscha}) is time dependent and singular when $r(x^+) = 0$. On similar lines, considering a generic null geodesic purely in the BTZ geometry and taking the appropriate plane wave limit, we obtain flat space. This is because BTZ is locally $AdS_3$ and plane wave limits of geodesics in maximally symmetric spaces always lead to flat space \cite{Blau:2002mw}. It is possible that these facts about Penrose limits in the geometry of black holes in $AdS$, that Penrose limits in $BTZ\times S^3\times M$ were not studied earlier. In this section, we show that the geometry of the Penrose limit of an in-falling geodesic in  $BTZ\times S^3\times M$  with angular momenta along $S^3$ is identical to the plane wave geometry obtained by examining the  Penrose limit of a time-like geodesic in   $AdS_3\times S^3 \times M$. Therefore, the plane wave spectra for the world sheet theories  in  these geometries are identical

\subsection{The Penrose limit of $BTZ\times S^3\times S^3\times S^1$}

In this section, we obtain the plane wave limit $BTZ\times S^3\times S^3\times S^1$ on in-falling geodesics. We choose to study the case of $M = S^3\times S^1$ with mixed Ramond-Ramond and NS fluxes with angular momentum on both $S^3$'s. Studying this general case enables the possibility of taking limits so that we can obtain a situation with either pure Ramond-Ramond flux or only pure NS flux, and with angular momentum only on one of the $S^3$. We will compare our result for the plane wave geometry with the plane wave limit of  $AdS_3\times S^3 \times S^3\times S^1$   with mixed fluxes obtained in 
\cite{Dei:2018yth}. The  gravity background of $BTZ\times S^3\times S^3\times S^1$  is given by 
\begin{equation}\label{form of metric}
ds^2 = G_{\mu\nu} dx^{\mu} dx^{\nu} = R^2 ds^2_{BTZ} + R_{1}^2 ds^2_{S_1^3} + R_{2}^2 ds^2_{S_2^3} + ds^2_{S^1},
\end{equation}
where 
\begin{align}\label{explicit form of metric of BTZS3S3S1}
  &  ds^2_{BTZ}  = -\left(\rho^2- M^2\right) d\tau^2 + \frac{d\rho^2}{\left(\rho^2- M^2\right)} + \rho^2 dx^2, \quad\quad\quad\quad \\
  &  ds^2_{S^3} = d\psi_1^2 + \sin^2{\psi_1} \ d\Omega^2_2, \qquad
    ds^2_{\tilde S^3} = d\psi_2^2 + \sin^2{\psi_2} \ d\tilde{\Omega}^2_2, \qquad
    ds^2_{S^1} = d{\tilde \Omega_1^2}.\nonumber
\end{align}
$d \Omega_2^2, d \tilde \Omega_2^2$  are metrics on unit 2-spheres and 
$\Omega_1$ is the angle on the circle $S^1$. 
Here, the radii $R$, $R_1$, and $R_2$ must satisfy the following condition to ensure the background solves the consistency condition of string theory in  10 dimensions. This leads to the constraint  \cite{Elitzur:1998mm}. 
\begin{equation}\label{Radius relation for BTZ_S3_S3_S1} 
 \frac{R^2}{R_1^2} + \frac{R^2}{R_2^2} =1.
\end{equation}
Following \cite{Dei:2018yth}, we parametrise this relation as
\begin{equation}\label{parametrization of Radius relation for BTZ_S3_S3_S1}
\frac{R^2}{R_1^2} = \alpha \equiv \cos^2 \varphi, \quad \frac{R^2}{R_2^2} = 1 - \alpha \equiv \sin^2 \varphi.
\end{equation}
The solution is supported by both  NS fluxes  and Ramond-Ramond fluxes, which are given by 
\begin{eqnarray} \label{fluxesdef}
& &H =  d B_{NS}=   2qR^2\lrt{\text{Vol}\lr{BTZ} + \frac{1}{\cos^2\varphi}\text{Vol}\lr{S^3}+ \frac{1}{\sin^2\varphi}\lr{\tilde S^3}}, 
\\ \nonumber
& &F =  2 \sqrt{ 1-q^2} R^2  \lrt{\text{Vol}\lr{BTZ} + \frac{1}{\cos^2\varphi}\text{Vol}\lr{S^3}+ \frac{1}{\sin^2\varphi}\lr{\tilde S^3}}, 
\end{eqnarray}
where  
\begin{eqnarray} 
& & {\rm Vol}(BTZ) =  \rho d \tau \wedge \rho \wedge  dx , \qquad \qquad 
{\rm Vol}(S^3 ) =  \sin^2 \psi_1 d\psi_1   \wedge {\rm Vol} S^2 ,  \nonumber \\
& & 
{\rm Vol}(\tilde S^3 ) =   \sin^2 \psi_2 d\psi_2   \wedge {\rm Vol} \tilde S^2 .
\end{eqnarray}
In (\ref{fluxesdef}), $0\leq q \leq 1$ is the parameter which dials between a background supported by 
pure NS flux and that supported by pure Ramond-Ramond flux. Consider now an in-falling geodesic in this background  whose trajectory obeys the equations 
\footnote{To relate factors:  $\alpha'$ in this paper  is equal to  $2\pi \alpha'$  in \cite{Dei:2018yth}.}
\begin{eqnarray}\label{isometry geodesic for BTZ_S3_S1}
   & & \dot{\tau} = \frac{E}{\left(\rho^2- M^2\right)},\quad \dot{x}= \frac{v}{\rho^2}, \\ \nonumber
    && \dot{\psi}_1 = \frac{ \alpha'}{R^2 }  \cos^2{\varphi}  J_1, \quad \dot{\psi}_2 = \frac{\alpha'}{R^2} 
    \sin^2\varphi  J_2 . 
\end{eqnarray}
The reason for the normalisations in the angular velocity is that the angular momentum from the Noether definition is defined as $J_1$. This can be seen as follows, 
from the Noether definition of the charge corresponding to shifts in $\psi_1$, WZW model on $S^3$ at level $k_1$, the angular momentum is given by 
\begin{eqnarray}
J_1 &=& \frac{k_1}{2\pi} \int_0^{2\pi} \dot \psi_1
= \frac{R_1^2}{ 2\pi \alpha' } \int_0^{2\pi } \dot \psi_1 , \\ \nonumber
&=&  \frac{R^2}{2\pi \alpha' \cos^2\varphi} \int_0^{2\pi } \dot \psi_1.
\end{eqnarray}
A very similar analysis explains the normalisation of the angular velocity $\dot \psi_2$ in (\ref{isometry geodesic for BTZ_S3_S1}). In the  2-spheres $S^2, \tilde S^2$  and the circle $S^1$, let us choose the geodesic to be at a fixed angle $\frac{\pi}{2}$ in all directions. Since the geodesic is null in the  $BTZ\times S^3\times S^3 \times S^1$ background, we have the relation
\begin{align}\label{radial equation in BTZS3S3S1}
  &  \dot{\rho}^2 + \frac{\alpha^{\prime 2} }{R^4 } J^2\lr{\rho^2-M^2} + v^2\rho^{-2}\lr{\rho^2-M^2}= E^2, \\ \nonumber
   &  J^2= ( J_1 \cos{\varphi})^2 + ( J_2 \sin{\varphi} )^2.
\end{align}
Note that comparing the equations for $\dot \tau, \dot x, \dot \rho$  in (\ref{isometry geodesic for BTZ_S3_S1}), (\ref{radial equation in BTZS3S3S1}) with that in (\ref{c1}, (\ref{c2}) and (\ref{radialeq}), we see that the trajectory is that of the in-falling geodesic in BTZ. We can make the identifications
\begin{eqnarray}\label{identifcations}
&& \kappa = \frac{\alpha' J }{R^2}  = \frac{J}{k} = \frac{ \Delta}{k}, \\ \nonumber
 && \kappa M \sinh\eta_1 \cosh\eta_2 = E, \qquad \kappa M \sinh\eta_2 \cosh \eta_1 = v.
\end{eqnarray}
In the first line in  (\ref{identifcations}), we have used the definition $k \kappa = \Delta$. Furthermore, the constraint in the second line of (\ref{radial equation in BTZS3S3S1}) allows the introduction of 2 angles defined as
\begin{eqnarray} \label{defangles}
\cos\omega = \frac{J_1}{J} \cos \varphi, \qquad \qquad 
\sin\omega = \frac{J_2}{J} \sin\varphi.
\end{eqnarray}

We perform the usual steps to take the Penrose limit along this geodesic. We first make the change of coordinates from $( \tau, \rho, x, \psi_1, \psi_2) \rightarrow ( \lambda , \xi,  \theta_0, \theta_1, \theta_2) $ which is given by 
\begin{align} \label{cooridnatetopp}
   &  \tau=\int_0^\lambda  \dot{\tau} \ d\lambda' - \frac{\xi}{E} + v\, \theta_0 + \frac{J_1\cos\varphi}{k}\theta_1 + \frac{J_2\sin\varphi}{k}\theta_2,  \nonumber \\
    &  \rho=\int_0^\lambda \dot{\rho}\ d\lambda',\qquad x=\int_0^\lambda \dot{x}\ d\lambda' + E \ \theta_0 ,\\ \nonumber
    &\psi_1= \cos{\varphi} \lrt{ \cos{\varphi} \frac{J_1}{k}\ \lambda + E\ \theta_1},\qquad \psi_2= \sin\varphi \lrt{ \sin\varphi \frac{J_2}{k}\ \lambda + E\ \theta_2}, \\ \nonumber
    &\qquad {\rm where} \;\; k = \frac{R^2}{\alpha'}.
\end{align}
We substitute this transformation in the metric (\ref{form of metric}), perform the following scaling 
\begin{eqnarray}\label{ppscaling}
& & \lambda \rightarrow \lambda,\qquad  \xi \rightarrow \frac{\xi}{R^2}, \qquad
\theta_i \rightarrow  \frac{\theta_i}{R},  \quad i = 0, 1, 2\\ \nonumber
& &\chi_a \rightarrow \frac{\pi}{2} +  \frac{\chi_a}{R_1}, \quad a\in  \Omega_2, \qquad 
\chi_a \rightarrow \frac{\pi}{2} + \frac{\chi_a}{R_2}, \quad  a \in \tilde\Omega_2, \\ \nonumber
&& \Omega^1 \rightarrow \frac{ \Omega^1}{R}.
\end{eqnarray}
Here $\chi_i$ refers to the angles of the 2-spheres $\Omega_1, \tilde \Omega_2$. We then  take the following limit
\begin{eqnarray} \label{largeR}
R\rightarrow \infty, \qquad {\rm with} \quad \frac{R}{R_1},\;  \frac{R}{R_2} \quad {\rm fixed}.
\end{eqnarray}
Substituting the transformation  (\ref{cooridnatetopp}) in the metric (\ref{form of metric}) and expanding in the large $R, R_1, R_2$ limit we obtain
\begin{eqnarray} \label{pplimit1}
   ds^2&=& 2d\lambda d\xi + \rho^2\lrt{E^2-\rho^{-2}v^2f}d\theta^2_0 + \lrt{E^2-\frac{\cos^2\varphi}{k^2}f J_1^2}d\theta_1^2+  \lrt{E^2-\frac{\sin^2\varphi}{k^2}f J_2^2}d\theta_2^2\nonumber\\
      &&- 2f\lrt{\lr{\frac{\cos\varphi}{k}vJ_1}\ d\theta_0d\theta_1 + \lr{\frac{\sin\varphi}{k}vJ_2}\ d\theta_0d\theta_2 + \lr{\frac{\sin\varphi\cos\varphi}{k^2}J_1J_2}\ d\theta_1d\theta_2 }\nonumber\\
      &&+\sin^2\lr{ {\frac{\cos^2\varphi}{k}J_1 \lambda}}\ ds^2\lr{\mathbb{R}^2} +    \sin^2\lr{{\frac{\sin^2\varphi}{k}J_2\lambda}}\ ds^2\lr{\tilde{\mathbb{R}}^2} + ds^2\lr{\mathbb{R}^1}.
\end{eqnarray}
Here $dS^2 ( \mathbb{R}^2)$  and similar expressions  refers to the Euclidean metric on $\mathbb{R}^2$ and 
\begin{eqnarray}
f = \rho^2 - M^2. 
\end{eqnarray}
This is the metric of the plane wave in Rosen coordinates. The transformation to the Brinkmann coordinates is easier once we diagonalise the metric in the directions $\theta_i$. We first decouple $\theta_1, \theta_2$ by the following orthogonal transformation
\begin{eqnarray} \label{thetatrans}
\tilde\theta _1 =  \frac{ J_1 \cos\varphi }{ J  } \theta_1 + \frac{J_2 \sin\varphi }{J } \theta_2 , 
\\  \nonumber
\tilde \theta_2 = - \frac{ J_1 \sin \varphi }{ J  } \theta_1 + \frac{J_2 \cos\varphi }{J } \theta_2.
\end{eqnarray}
This brings the metric (\ref{pplimit1})  to  the form 
\begin{align} \label{pplimit2}
   ds^2&= 2d\lambda d\xi + \rho^2\lrt{E^2-\rho^{-2}v^2f}d\theta^2_0 + \lrt{E^2- \frac{J^2}{k^2} f}d\tilde{\theta}_1^2 
    -2vf\frac{J}{k} d\theta_0d\tilde{\theta}_1 +E^2d\tilde{\theta}_2^2  \nonumber\\
     &+\sin^2\lr{ {\frac{\cos^2\varphi}{k}J_1\lambda}}\ ds^2\lr{\mathbb{R}^2} +    \sin^2\lr{{\frac{\sin^2\varphi}{k}J_2\lambda}}\ ds^2\lr{\tilde{\mathbb{R}}^2} + ds^2\lr{\mathbb{R}^1}.
\end{align}
We now have to decouple $\theta_0$ and $\tilde \theta_1$; let us write the metric in this subspace explicitly. 
\begin{eqnarray} \label{defarho}
d\tilde{s}^2 &=&  ( d\theta_0 , \; d\tilde \theta_1 ) \; A( \rho) \; 
\left( \begin{array}{c}
d\theta_0 \\ d\tilde\theta_1
\end{array}
\right), \\ \nonumber
     A(\rho) &=& \rho^2\begin{pmatrix}
         E^2-v^2& -v j \\
         -v j &- j^2 
     \end{pmatrix}
     + \begin{pmatrix}
       v^2M^2& v j M^2  \\
       v j M^2 &E^2 +  j^2 M^2
   \end{pmatrix} , \quad j \equiv \frac{J}{k}. \nonumber\\ \nonumber
   &\equiv&\rho^2 B + C
\end{eqnarray}
To bring the matrix $A$ to the diagonal form, we first perform an orthogonal transformation from the coordinates $( \theta_0, \tilde \theta_1)\rightarrow (\hat \theta_0, \hat \theta_1)$ to which $C$ is diagonal. This transformation is independent of $\rho$ and brings $C$ to the following form 
{\small
\begin{eqnarray} \label{diagonal}
& & O^T C O  = 
\left( 
\begin{array}{cc}
c_- & 0 \\
0 & c_+
\end{array}
\right) 
,  \\ \nonumber
& & c_\pm =  \frac{1}{2} \Big( E^2 +  M^2( j^2 + v^2) \pm  \sqrt{ \big[  E^2 +  M^2( j^2 + v^2)]^2 - 4 v^2 E^2 M^2 \big] }
\Big).
\end{eqnarray}
}
When $v\neq 0$,  none of the eigenvalues are zero. In this basis, we  perform the re-scaling 
\begin{eqnarray} \label{thetascale}
(\hat\theta_0, \hat \theta_1) \rightarrow  \Big(\frac{ \check\theta_0}{\sqrt{c_-} }, \frac{\check\theta_1}{\sqrt{c_+} } \Big).
\end{eqnarray}
In the coordinates $(\check\theta_0, \check\theta_1)$, the matrix $C \rightarrow I $ has become the identity, and $B$ has become  $\check B$, a symmetric matrix. 
We now perform the final orthogonal transformation to make the matrix  $\check B$ a diagonal matrix. Note that this transformation will retain $C$ as identity. 
This sequence of coordinate transformations, which are independent of $\rho$, has the following effect on the matrix 
$A(\rho)$
{\small
\begin{eqnarray} \label{lamdbdapm}
&&A(\rho) = B\rho^2 +  C \rightarrow 
\left( 
\begin{array}{cc}
          1 + \lambda_+ \rho^2& 0\\
          0&1 + \lambda_-\rho^2
          \end{array}
     \right), \\ \nonumber
   &&   \text{where}\quad \lambda_{\pm} =\frac{E^2 + j^2 M^2 - v^2 \pm \sqrt{(E^2 + j^2 M^2)^2 - 
  2 (E - j M)(E + j M) v^2 + v^4}}{2 M^2 v^2}.
\end{eqnarray}
}
Therefore from the metric in (\ref{pplimit2}), we have arrived at the following metric
 \begin{align} \label{pplimit3}
   ds^2&= 2d\lambda d\xi +  \lr{1 + \lambda_+\rho^2}d\theta^2 +  \lr{1 + \lambda_-\rho^2}d\tilde{\theta}^2+\sin^2\lr{ {\frac{\cos^2\varphi}{k}J_1\lambda}}\ ds^2\lr{\mathbb{R}^2}\nonumber\\
   &\quad\quad +    \sin^2\lr{{\frac{\sin^2\varphi}{k}J_2\lambda}}\ ds^2\lr{\tilde{\mathbb{R}}^2} + 
   E^2 d\tilde \theta_2^2 + ds^2\lr{\mathbb{R}^1}.
\end{align}
Here we have labelled the final decoupled co-ordinates as $(\theta, \tilde \theta)$. The plane wave metric in (\ref{pplimit2})  is in the Rosen form, and since it is diagonal, we can go over to the Brinkmann co-ordinates. Let us recall the general procedure. Consider a metric of the form 
\begin{eqnarray}
ds^2 = 2 d\lambda d\xi  + \sum_{i=1}^8 C_i^2 (dy^i)^2.
\end{eqnarray}
  We go to Brinkmann coordinates by the following coordinate transformation
\begin{equation}\label{Brinmann coordinate transformation}
    \lambda=x^+ ,\quad \xi=x^- +\frac{1}{2}\sum\limits_{i} \frac{1}{C_i}\lr{\frac{\partial C_i}{\partial x^+}}x^ix^i,\quad y^i= C^{-1}_ix^i.
\end{equation}
Then the metric becomes 
\begin{align}
    ds^2&= 2dx^+dx^- + \lrt{\sum\limits_{i} {\frac{C^{''}(x^+)}{C_i} x^ix^i}}\lr{dx^+}^2 + \sum\limits_{i}dx^idx^i\nonumber\\
    &=2dx^+dx^- + \sum\limits_{i}  \lr{ M_{ii}x^ix^i}\lr{dx^+}^2 + \sum\limits_{i}dx^idx^i\nonumber.
\end{align}
To evaluate the matrix $M$, we first write  the derivative with respect to  $\lambda$ in terms of the derivative 
with respect to $\rho$  as,
\begin{align}\label{derivative w.r.t u in terms of rho}
    \frac{d^2}{d\lambda^2}= \frac{d}{d\lambda}\lr{\dot{\rho}\frac{d}{d\rho}}= \dot{\rho}^2\frac{d^2}{dr^2} + \dot{\rho}\lr{\frac{d\dot{\rho}}{d\rho}}\frac{d}{d\rho}\ .
\end{align}
Carrying out this procedure on the metric in (\ref{pplimit3}), we obtain 
\begin{eqnarray}
    ds^2&=& 2dx^+dx^- - \lrt{ j^2(x_1^2 + x_2^2) + \frac{\cos^4{\varphi}}{k^2}J_1^2(x_3^2+ x_4^2) +  \frac{\sin^4{\varphi}}{k^2}J_2^2(x_5^2+ x_6^2)} (dx^+)^2   \nonumber \\
   && \qquad  + \sum_{i = 1}^8dx^idx^i . 
\end{eqnarray}
This is a homogeneous plane wave metric with masses in $6$ directions.  It is important to note that the masses in the directions $x_1,x_2$ result because of the precise form of the metric in $\theta, \tilde \theta$ directions in (\ref{pplimit3}) and the values of $\lambda_\pm$ in (\ref{lamdbdapm}). We can express $J_1, J_2$ in terms of the angles introduced in (\ref{defangles}), this results in 
\begin{eqnarray} \label{finalpplimit}
    ds^2 &=& 2dx^+dx^- - j^2  \lrt{ (x_1^2 + x_2^2) + \cos^2 {\varphi} \cos^2 \omega  (x_3^2+ x_4^2) +  
    \sin^2 \varphi \sin^2 \omega (x_5^2+ x_6^2)} (dx^+)^2  \nonumber \\
    &&  \qquad\qquad + \sum_{i = 1}^8dx^idx^i .
\end{eqnarray}
We can now compare this metric to that obtained by taking the plane wave limit of the time-like geodesic at the origin in $AdS_3\times S^3\times S^3 \times S^1$ with angular momentum along $S^1$'s in the 3-spheres obtained in \cite{Dei:2018yth} \footnote{Please see equation (2.20), (2.21), (2.22) of \cite{Dei:2018yth}. Note that the overall  $j$ in (\ref{finalpplimit}) can be scaled away by redefining $x^+, x^-$. }.
We see that the mass matrix precisely agrees. Let us now examine the limiting forms of the background fluxes. We take limits of each component of the fluxes in (\ref{fluxesdef}), by performing the coordinate transformation in (\ref{cooridnatetopp}) and then performing the scaling  (\ref{ppscaling}) and taking the limit (\ref{largeR}). 
\begin{eqnarray}
 &&  R^2\ \text{Vol}(BTZ)= R^2 \rho\ d\tau\wedge d\rho\wedge dx \\ \nonumber
   &&\quad =R^2\rho\ \lr{\dot{\tau} \ d\lambda - \frac{d\xi}{E} + vd\theta_0 + \frac{J_1\cos^2\varphi}{k}d\theta_1 + \frac{J_2\sin^2\varphi}{k}d\theta_2}\wedge \dot{\rho}\ d\lambda\ \wedge \lr{\dot{x}d\lambda + E\ d\theta_0}.
   \end{eqnarray}
   Now performing the scaling (\ref{ppscaling}), taking the large $R$ limit 
   and using the transformation (\ref{thetatrans}),  we obtain 
   \begin{eqnarray}
     \lim_{R\rightarrow \infty} R^2\ \text{Vol}(BTZ)
   &=&\rho\ \dot{\rho}\ E  \lr{\frac{J_1\cos^2\varphi}{k}d\theta_1 + \frac{J_2\sin^2\varphi}{k}d\theta_2}\wedge d\lambda\wedge d\theta_0\nonumber\\
   &=&jE \rho\dot{\rho}\ d\tilde{\theta}_1\wedge d\lambda \wedge d\theta_0.
   \end{eqnarray}
   To proceed further, observe that, 
   \begin{eqnarray} 
  E \rho \dot \rho  =  \Big( {\rm det }A(\rho)  \Big)^{\frac{1}{2}}   = v EM \sqrt{( 1+ \lambda_+ \rho^2)( 1+ \lambda_- \rho^2)}.
   \end{eqnarray}
   This, together with the fact scaling of the co-ordinates in (\ref{thetascale}), implies that we can write 
   \begin{eqnarray} \label{pplimivolbtz}
    \lim_{R\rightarrow \infty} R^2\ \text{Vol}(BTZ)
   &=&j\ d\lambda\wedge\lr{\sqrt{1 + \lambda_+ \rho^2}\ d\theta}\wedge \lr{\sqrt{1 + \lambda_-\rho^2}\ d\tilde{\theta}}\nonumber\\
   &=&j\ dx^+\wedge dx_1 \wedge dx_2.
\end{eqnarray}
In the last line, we have used  the  transformation to the Brinkmann coordinates 
(\ref{Brinmann coordinate transformation}). 
Now, let us examine the fluxes on $S^3$
\begin{eqnarray}
 &&  \frac{R^2}{\cos^2 \varphi}  \text{Vol}(S_1^3)
   = \frac{R^2}{\cos^2 \varphi}  \sin^2{\psi_1}\ d\phi\wedge d\text{Vol}(S^2) \\ \nonumber
    &&\quad =  \frac{R^2}{\cos^2 \varphi}  
    \sin^2\lr{\frac{\cos^2{\varphi}}{k }J_1\lambda + E \cos \varphi \; \theta_1}\ \lr{\frac{\cos^2{\varphi}}{k}J_1d\lambda + \cos\varphi E \; d\theta_1}\wedge d\text{Vol}(S^2).
    \end{eqnarray}
    We scale the co-ordinates as in (\ref{ppscaling}) and take the $R\rightarrow\infty$ limit. 
    \begin{eqnarray} \label{pplimivols31}
    \lim_{R\rightarrow\infty}    \frac{R^2}{\cos^2 \varphi}  \text{Vol}(S_1^3)
    &=&\frac{\cos^2\varphi}{k }J_1 
    \sin^2\lr{\frac{\cos^2\varphi}{k}J_1 \lambda}d\lambda\wedge d\text{Vol}(\mathbb{R}^2) \nonumber\\
    &=& j \cos \varphi  \cos dx^+\wedge dx_3\wedge dx_4.
\end{eqnarray}
In the last line, we have used the definition (\ref{defangles}) to relate $J_1$ to $\cos \omega$. Proceeding along similar lines, we obtain the limit 
\begin{eqnarray}  \label{pplimivols32}
  \lim_{R\rightarrow\infty}  \frac{R^2}{\sin^2 \varphi} \ \text{Vol}(S_2^3)&=& j \sin\varphi \sin \omega dx^+\wedge dx_5 
  \wedge dx_6.
  \end{eqnarray}
  From the limit in (\ref{pplimivolbtz}), (\ref{pplimivols31}) and (\ref{pplimivols32}), we see that in the plane wave limit, the fluxes are given by 
  \begin{eqnarray} \label{finalfluxlimit}
      H&=& 2q\ j\ dx^+ \wedge\lrt{\  dx_1 \wedge dx_2 + j\cos\omega\cos\varphi \ dx_3 \wedge dx_4 + j\sin\omega \sin\varphi \ dx_5 \wedge dx_6 }, \\ \nonumber
      F &=& 2 \sqrt{1-q^2} 
      \ j\ dx^+ \wedge\lrt{\ dx_1 \wedge dx_2 + j\cos\omega\cos\varphi \ dx_3 \wedge dx_4 + j\sin\omega \sin\varphi \ dx_5 \wedge dx_6 }.
  \end{eqnarray}

  The  analysis  of the Penrose limit 
  can be repeated for the case $v=0$ for which the metric $A(\rho)$ in (\ref{defarho}) is diagonal \footnote{The analysis is simpler for $v=0$ and it does not involve the diagonalization in  (\ref{diagonal}). }
  It can be shown that we obtain the identical limit for the metric and fluxes. We see that the metric (\ref{finalpplimit}) and the fluxes (\ref{finalfluxlimit}) in the plane wave limit along the in-falling geodesic on $BTZ\times S^3\times S^3\times S^1$ and the  Penrose limit along the time-like geodesic at the origin in $ AdS_3\times S^3\times S^1$  obtained in \cite{Dei:2018yth} precisely agree. This implies that the spectrum of excitations of string theory in this plane wave background is identical to that in  $ AdS_3\times S^3\times S^1$. We will not repeat the analysis since it is identical to that done in \cite{Dei:2018yth}.  
  In the case of the  plane waves in $BTZ$, the light cone Hamiltonian is the charge 
  \begin{eqnarray}
  {\cal H} &=& \Delta -\cos \varphi \cos \omega J_1 - \sin \varphi \sin \omega J_2, \\ \nonumber
  &=& k \sqrt{ -Q_L \cdot Q_L }   -\cos \varphi \cos \omega J_1 - \sin \varphi \sin \omega J_2\ .
  \end{eqnarray}
  The second line arises because of the identification of the charge $\Delta = k \kappa$ along time-like geodesics in $AdS_3$  with the Casimir of the $SL(2, R)$ sigma model in the BTZ geometry, as shown for example in  (\ref{casimir1}). It should be emphasised that the resultant geometry after the Penrose limit is independent of the initial conditions of the in-falling geodesic and therefore true for any geodesic in-falling into the horizon of the BTZ geometry. The plane wave limit of the spectrum of the WZW model on $AdS_3\times S^3\times S^3\times S^1$ also yields the same dispersion relation obtained by quantising the string theory on the plane wave geometry \cite{Dei:2018yth}. Since we have demonstrated that the geometry obtained in the region of the in-falling geodesic on $BTZ \times S^3\times S^3 \times S^1$ is identical to that of the plane wave limit in the $AdS_3$, it must be the case that there exists a limit of the spectrum of the WZW model on $BTZ \times S^3\times S^3 \times S^1$ which results in identical spectrum as in case of $AdS_3$. The approach would involve expanding the highest weight state in a direction which is boosted by the parameters $\eta_1, \eta_2$. This is because we showed in section \ref{sectionvel} that the charges of the time-like geodesics in $AdS_3$, which are semi-classical states in the WZW model, are related to those of the in-falling geodesic in BTZ by an $SL(2, R)$ boost.  These charges are time-like in the $SL(2, R)$ norm, as can be seen in (\ref{casimir1}) and (\ref{casimir2}). Therefore, it suggests that expanding the spectrum around the background of a state with a large time like $SL(2, R)$ charge in the BTZ sigma model results in a plane wave spectrum. It would be interesting to do this explicitly.

\section{In-falling strings as local quenches}\label{sec4}

In this section, we show that the holographic duals of in-falling geodesics, circular strings, as well as magnons in the BTZ are local time-dependent quenches. These quenches are more general compared to those studied earlier in the literature in the $AdS_3/CFT_2$ context
\cite{Asplund:2011cq,Nozaki:2013wia,Nozaki:2014hna,Caputa:2014vaa,Asplund:2014coa,Caputa:2014eta,David:2016pzn}. We will see that the quenches here carry R-charges, marginal operators acquire non-trivial expectation values, and the left-moving and right-moving pulses of the quenches are not symmetric.  In spite of the large amount of literature on local quenches, especially in the context of $AdS_3/CFT_2$, such quenches have not been studied earlier. Let us first recall the $AdS_3/CFT_2$ dictionary for a particle  of mass $m$ at the origin of  global $AdS_3$
\cite{Nozaki:2013wia}. The $CFT_2$ is on the cylinder parameterised by $(t, \phi)$ with $\phi \sim \phi +2\pi$. At finite gravitational coupling $G_N$ particle at the origin of $AdS_3$ backreacts and induces a conical defect. This can be seen by considering Einstein's equation \footnote{We have used the $\hat R$ to distinguish the Riemann curvature from the radius of curvature $R$ of $AdS_3$. } 
\begin{eqnarray}
\hat R_{\mu\nu} - \frac{1}{2} g_{\mu\nu} \hat R  - \frac{g_{\mu\nu} }{R^2} = 8\pi  G_{N}  T_{\mu\nu},
\end{eqnarray}
where  the only non-zero component of the  stress tensor is given by 
\begin{eqnarray}
T^{tt} = \frac{m}{r} \delta (r) \delta( \phi). 
\end{eqnarray}
We solve Einstein's equation and obtain the  backreacted metric, which is the metric of the conical defect  
\begin{eqnarray} \label{conical}
ds^2 = - ( r^2 + R^2 - \tilde m ) dt^2 + \frac{R^2}{r^2 + R^2 - \tilde m } dr^2 + r^2 d\phi^2,
\end{eqnarray}
with 
\begin{eqnarray}
\tilde m = 8 G_N R^2 m. 
\end{eqnarray}
We can read out the stress tensor from the back-reacted geometry using the Fefferman-Graham expansion. This allows us to conclude that the state in the $CFT_2$ is created by the insertion of a primary operator  ${\cal O}$ at $t\rightarrow -\infty$ on the cylinder. The dimension of the operator and the mass of the particle are related by 
\begin{eqnarray} \label{mass-dim}
m  = \frac{\Delta_{\cal O} }{R}, \qquad \qquad \tilde m = 8G_N R \Delta_{\cal O }  = \frac{12 \Delta_{\cal O}  R^2}{c}.
\end{eqnarray}
To obtain the last equality, we have used the Brown-Henneaux formula \cite{Brown:1986nw}
\begin{eqnarray} \label{brownhen}
\frac{1}{G_N} = \frac{2 c}{3R}.
\end{eqnarray}
Thus, the state in the CFT can be written as 
\begin{eqnarray}
|{\cal O } \rangle =\lim_{t\rightarrow -\infty}  {\cal O } ( t, 0 ) |0 \rangle. 
\end{eqnarray}

When this particle is in-falling in the BTZ geometry, it has been shown that the dual description of the back-reacted geometry is that of a local quench. The back-reacted geometry can be obtained by using the coordinate transformation (\ref{coordinate transformation for in-falling geodesic}) on the conical defect geometry (\ref{conical}). The state in the CFT  is described by the density matrix 
\begin{eqnarray}
& &\rho_\epsilon  = {\cal N } e^{- i Ht} \Big( e^{-\epsilon H} {\cal O }(0, 0 ) e^{\epsilon H} \Big ) \rho_{\beta} 
\Big( e^{-\epsilon H} {\cal O }^\dagger (0, 0 ) e^{\epsilon H} \Big ) e^{i H t}, \nonumber \\ 
& & \rho_\beta = e^{-\beta H}.  
\end{eqnarray}
Here $\rho_\beta$ is the thermal density matrix of the dual CFT and ${\cal O}(0, 0 )$ is the conformal primary field generating an excitation at $t=0$, placed at the origin in the spatial direction in the thermal CFT. ${\cal N}$ is introduced to normalise the density matrix. The expectation value of the stress tensor on this state is time-dependent. It starts out at $t=0$ as a single pulse at the origin, splits into $2$ symmetric left and right-moving pulses. These pulses travel at the speed of light. The height of the pulse is proportional to the operator dimension and therefore related to the mass of the in-falling particle in the dual geometry. The  variable $\epsilon$ is the width of the pulse and can be shown to be related to the $\eta_1$ of the map 
(\ref{coordinate transformation for in-falling geodesic}), or to the radial position at which the particle is released using the initial conditions (\ref{initialcond}).  Our subsequent discussion will review and generalise these facts about local quenches. From the results in sections \ref{sec2}, \ref{sec3}, we see that the in-falling geodesics, circular strings, and magnons follow the same trajectory in $BTZ$ as massive geodesics. However, their mass arises due to the angular momenta in the compact  $S^3$.  They also have velocity in the longitudinal direction. In this section, we show that the dual description of these in-falling classical solutions is
local quenches as described earlier, in addition, they carry expectation values of $R$-charges due to the angular momentum in the $S^3$ and expectation values of scalars that arise from the dimensional reduction on $S^3$. 
We show that these quenches are not symmetric pulses moving on the left and right light cones in the $CFT_2$, the parameter $\eta_2$  in the map (\ref{coordinate transformation for boosted in-falling geodesic}), which turns on velocity of the in falling particle in the BTZ renders the width as well as the height of the left moving and right moving pulse asymmetric. The energy density profile of this quench is shown in Figure \ref{AsymmetricQuenchStressTensor}. In the dual thermal CFT, the dual state is the density matrix 
\begin{eqnarray}
\rho = {\cal N } e^{- i  Ht }  {\cal O }( -i \epsilon_1,  i \epsilon_2 )  \, 
 \rho_{\beta}  \, 
   {\cal O }^\dagger (i \epsilon_1 , -i \epsilon_2  )  e^{i H t}.
\end{eqnarray}
The $\epsilon_i$ are parameters which regularise the local quench,  the quenches studied earlier in \cite{Caputa:2014eta,David:2016pzn}  considered the situation with  $\epsilon_1 = \epsilon_2$. $\rho_\beta$ is the thermal density matrix and ${\cal N}$ is the normalisation chosen to ensure ${\rm Tr}(\rho^2) = 1$. We will show that the parameters $\epsilon_1, \epsilon_2$ in the CFT  are related to the parameters $\eta_1, \eta_2$ of the maps of the in-falling particle (\ref{coordinate transformation for boosted in-falling geodesic}). As far as we are aware,  our study is the first instance of asymmetric quenches and that of 
embedding local quenches fully in the  $AdS_3\times S^3 \times M$ background.

 To proceed, we first study the particle localised in the centre of $AdS_3$ carrying angular momentum along the $S^3$ and show that the dual description is that of a state in the CFT carrying  $R$ charge and expectation value of a marginal operator dual to the scalars that arise from the compactification on $S^3$.  We then generalise this observation to circular strings and giant magnons. This is done by obtaining the back-reacted geometry of these solutions and reading off the relevant $R$-charges and the expectation values. We then use the maps (\ref{coordinate transformation for in-falling geodesic}) and (\ref{coordinate transformation for boosted in-falling geodesic}) to obtain the back-reacted geometry of in-falling 
geodesics, circular strings and magnons. This allows us to show that their dual description in the thermal CFT is as local quenches carrying R-charges and expectation values of marginal operators.

\subsection{CFT duals of geodesics solutions in $AdS_3\times S^3\times M$}
\label{seccftdualads3}

We begin by considering a point-like string solution localised at the origin of $AdS_3$ and rotating along one of the angular directions of $S^3$.  This is a BPS configuration for all $M$ being $T^4, K^3$ or $S^3 \times S^1$. The world sheet couples to gravity via the action 
\begin{eqnarray} \label{swscoupling}
S_P =  \frac{k}{4 \pi } \int d^{10} x  \sqrt{G} 
 \int d^2 \sigma  \Big[  G_{\mu \nu } \partial_a x^\mu \partial^a x^\nu  \Big] \frac{\delta(  x^\mu - \bar x^\mu ( \sigma) ) }
{\sqrt{ G} }. 
\end{eqnarray}
Here $\mu, \nu $  take values in $\{ 0, 1, \cdots 9 \}$. 
To be specific, let us take the target space metric to be $AdS_3\times S^3 \times T^4$, the discussion 
can be easily generalised to the other situations. 
\begin{eqnarray}
ds^2 = G_{\mu\nu} dx^\mu dx^\nu =  ds^2_{AdS_3}  + R^2 ds_3^2 + ds_{M}^2.
\end{eqnarray}
where the metric on the $AdS_3$ and $S^3$  is given in (\ref{ads3}), (\ref{mets3}).  
The trajectory of the particle at the origin and spinning  along the angle $\phi_2$ of the sphere is given by 
\begin{eqnarray}
&&t( \sigma^0, \sigma^1) = \kappa \sigma^0, \qquad r ( \sigma^0, \sigma^1) = 0, \qquad 
\phi( \sigma^0, \sigma^1)  =0, \\ \nonumber
&&\psi( \sigma^0, \sigma^1) = 0, \qquad \phi_1( \sigma^0, \sigma^1) =0, \qquad 
\phi_2( \sigma^0, \sigma^1) = \frac{J}{k} \sigma^0 , \\ \nonumber
&& X( \sigma^0, \sigma^1)|_M = x_M = {\rm constant}.
\end{eqnarray}
The Virasoro condition, or the geodesic being massless, leads to the equality
\begin{eqnarray}
\Delta = J, \qquad \qquad\qquad \Delta = k \kappa\ . 
\end{eqnarray}
The stress tensor sourced by this particle is obtained by the definition
\begin{eqnarray}
T_{\mu \nu} =  - \frac{2}{ \sqrt{G} } \frac{\delta  S_P}{\delta G^{\mu\nu} }. 
\end{eqnarray}
The non-zero components of the 10-dimensional stress tensor are given by 
\begin{eqnarray} \label{10dstress}
T^{t\,t} =  \frac{\Delta}{R} \frac{ \delta(r)\delta( \phi) }{r}
\frac{  \delta(\psi) \delta( \phi_1) \delta ( \phi_2 - t) }{\sqrt{g_{S^3} }}
\frac{  \delta^4( X_M - x_M) }{\sqrt{G_M}} ,  \\ \nonumber
T^{t\,\phi_2} =  \frac{\Delta}{R} \frac{ \delta(r)\delta( \phi) }{r}
\frac{  \delta(\psi) \delta( \phi_1) \delta ( \phi_2 - t) }{\sqrt{g_{S^3} }}
\frac{  \delta^4( X_M - x_M) }{\sqrt{G_M}}, \\ \nonumber
T^{\phi_2\,\phi_2} = \frac{\Delta}{R} \frac{ \delta(r)\delta( \phi) }{r}
\frac{  \delta(\psi) \delta( \phi_1) \delta ( \phi_2 - t) }{\sqrt{g_{S^3} }}
\frac{  \delta^4( X_M - x_M) }{\sqrt{G_M} }.
\end{eqnarray}
To arrive at these expressions, we have performed the integral over the world sheet coordinates and used $\Delta =J$. Note that all the non-zero components are proportional to $\Delta$.

Let us dimensionally reduce these sources on $AdS_3\times S^3$ and focus on the lowest Kaluza-Klein mode on the 
sphere $S^3$. 
This can be done by integrating out the coordinates on $S^3\times M $. 
Therefore,  dimensional reduction  of the stress tensor components  in (\ref{10dstress}) 
  on $S^3\times M$, we obtain the following sources
  \begin{eqnarray} \label{sources}
  \hat T^{t\,t} =  \frac{\Delta}{R} \frac{ \delta(r)\delta( \phi) }{r} ,  \\ \nonumber
\hat J^t \equiv T^{t\,\phi_2} =    \frac{\Delta}{R} \frac{ \delta(r)\delta( \phi) }{r}, \\ \nonumber
\hat J_{\psi} \equiv T^{\phi_2,\phi_2} = \frac{\Delta}{R} \frac{ \delta(r)\delta( \phi) }{r}.
  \end{eqnarray}
The energy density $T^{tt}$ is a source for the Einstein's equations in $AdS_3$,  the 
stress tensor component $T^{t\, \phi_2}$ is a source for a $U(1)$ gauge field $A_\mu$  in $AdS_3$. 
Since this component of the stress tensor is the charge density for the $U(1)_R$ isometry of the $S^3$, it should 
source the gauge field corresponding to this symmetry.  
In $AdS_3$ the R-symmetry gauge field obeys the Chern-Simons equations of motion \cite{deBoer:1998kjm,David:1999nr} \footnote{See \cite{Deger:2014ofa,Samtleben:2019zrh} for the action obtained after dimensional reduction 
of gauged supergravity from $6$ dimensions on $AdS_3\times S^3$.} ,
 therefore $T^{t\, \phi_2}$  sources 
a Chern-Simons gauge field. 
The component $T^{\phi_2,\phi_2} $ sources a scalar $\psi$  in $AdS_3$. 
 The scalar field $\psi$ we choose to focus on is the dilaton in $6$ dimensions, which is massless when compactified
 on $AdS_3\times S^3$, see table (6.15)  of \cite{David:2002wn}.  The stress tensor $T^{\phi_2,\phi_2} $  couples to the $G_{\phi_1\phi_1}$ component of the metric in the string frame; therefore, it would couple to the dilaton in $6$ dimensions in the Einstein frame. 
 From these considerations, the relevant  dimensionally reduced effective action is given by 
     \begin{eqnarray} \label{3daction}
   S = \frac{1}{16\pi G_N} \int d^3 x \sqrt{g} \left( \hat R +  \frac{2}{R^2}    
    +\frac{R}{2}\frac{\epsilon^{\mu\nu\rho}}{\sqrt{g} }  A_\mu \partial_\nu A_\rho
   - \partial_\mu \psi\partial^\mu \psi
    \right). 
  \end{eqnarray}
  Here $\mu, \nu, \rho$ take values along the $AdS_3$ directions and $\epsilon^{tr\phi} = 1$. 
  The factor of $R$ in front of the Chern-Simons is required since the level of the Chern-Simons is proportional to the central charge by supersymmetry. The central charge is given by the Brown-Henneaux formula (\ref{brownhen}). 
  Therefore we have 
  \begin{eqnarray}
  k_{{\rm Chern\;Simons}} = \frac{c}{6} = \frac{R}{4 G_N}. 
  \end{eqnarray}
  The sources in (\ref{sources}) couple to the fields  by the following interaction term
  \begin{eqnarray} \label{sint}
  S_{\rm int} = \int  d^3 x  \sqrt{g} \left(  
  - \frac{1}{2} T_{\mu \nu } \delta g^{\mu\nu} -  J_\mu \delta A^\mu - \hat J_{\psi} \delta \psi
  \right). 
  \end{eqnarray}

  This is obtained by dimensional reduction of the coupling of the relevant fluctuations to the higher-dimensional metric. Considering the action in (\ref{3daction}) and the sources in (\ref{sint}), the dimensionally reduced equations of motion in $AdS_3$  become 
  \begin{eqnarray}\label{3deom}
  \hat R_{\mu\nu} - \frac{1}{2} g_{\mu\nu}  \hat R -\frac{ g_{\mu\nu}}{R^2} = 8\pi G_N \hat T_{\mu\nu} , \\ \nonumber
  \frac{R}{\sqrt{g} } \epsilon^{\mu \nu \rho} \partial_\nu A_\rho   = 16 \pi G_N \hat J^\nu, \qquad \\ \nonumber
  \frac{1}{\sqrt{g} } \partial_\mu \Big( \sqrt{g} g^{\mu \nu} \partial_\mu \psi \Big)   =    8 \pi G_N \hat J_\psi.
  \end{eqnarray}
 $G_N$ is Newton's constant in 3 dimensions. As we are interested in the backreacted solution to the leading order in $G_N$, we can ignore the stress tensor of the scalar in Einstein's equation. From the equations of motion of the scalar, we see that these fields are of the order  $G_N$, and therefore they contribute at $G_N^2$ to the stress tensors. 
 The Chern-Simons field does not contribute to the stress tensor. 
 Therefore, the equations of motion are that for the metric,  the 
 Chern-Simons $U(1)$ field and a minimally coupled massless scalar $\psi$ in the background of $AdS_3$ with their corresponding sources at the origin. The AdS/CFT duality relates metric fluctuations to the conserved stress tensor of the CFT and the gauge field in the bulk to a conserved $U(1)$ R current in the CFT. 
A minimally coupled scalar field  of mass $m$  in the bulk is dual to a primary operator  $\Psi$, of conformal dimension $\Delta$, which is given by 
\begin{eqnarray}
\Delta_\Psi = 1 + \sqrt{ 1+ m^2R^2}. 
\end{eqnarray}
Here we have specialised to the  $AdS_3/CFT_2$  case. Thus, the massless scalar $\psi$ in the bulk is dual to a marginal scalar operator $\Psi$ with conformal dimension $\Delta_\Psi= 2$. Using AdS/CFT correspondence, we can read out the expectation value of the stress tensor and the $U(1)$ current. We will also show that the backreacted solution obtained by solving the equations in (\ref{3deom}) turns on the expectation value of the boundary stress tensor, a background  $U(1)$ charge, as well as an expectation value of the operator $\Psi$ in the boundary CFT. We then show that such expectation values are non-zero for a state in the CFT created by an operator ${\cal O}$ which has conformal dimension and $U(1)$ charge equal to $\Delta$ \footnote{$\Delta$ is the strength of the sources in (\ref{sources}).}. In addition to this, the operator must have a non-trivial $3$-point function with the marginal operator $\Psi$ equal to 
$\Delta$. 

 \subsubsection*{The backreacted solution of the spinning geodesic}

The solution to Einstein's equation in (\ref{3deom}) is known  and is given by the conical defect metric 
  \begin{eqnarray} \label{conical1}
ds^2 = - ( r^2 + R^2 - \tilde m ) dt^2 + \frac{R^2}{r^2 + R^2 - \tilde m } dr^2 + r^2 d\phi^2,
\end{eqnarray}
with 
\begin{eqnarray}
\tilde m = 8 G_N R \Delta.
\end{eqnarray}
Though this is an exact solution to Einstein's equation, as written in  (\ref{3deom}),  we have neglected the stress tensor of the scalar and the photon in the equation; therefore, the above solution is the leading order solution in $G_N$. We now solve the equations of motion for the scalar field in (\ref{3deom}). It is clear that to obtain the scalar field at the leading order in $G_N$, we can assume the background metric to be $AdS_3$. Since the source is radially symmetric,  the scalar field only depends on the coordinate $r$. The Laplacian on $AdS_3$ reduces to 
\begin{eqnarray}
\frac{1}{rR} \partial_r \Big[ r R ( r^2+ R^2) \partial_r \psi \Big]=  8\pi  G_N \frac{\Delta}{R r} \delta ( r) \delta( \phi). 
\end{eqnarray}
Solving for $\psi$, we obtain
\begin{eqnarray}
\psi =   C + \frac{2 \Delta G_N}{ R} \log \left( \frac{ r^2}{ r^2 + R^2} \right). 
\end{eqnarray}
We can set $C=0$ by requiring the solution to be normalizable at $r\rightarrow \infty$. 
Thus we have 
\begin{eqnarray} \label{solpsi}
\psi &=&   \frac{2\Delta G_N }{ R} \log \left( \frac{ r^2}{ r^2 + R^2} \right)  \\ \nonumber
&=& - \frac{2\Delta G_N}{R} \frac{R^2}{r^2} + \cdots
\end{eqnarray}
Observe that the scalar field falls off at the boundary as $\frac{R^2}{r^2}$, which is the expected fall-off behaviour for the expectation value of the dual operator  $\Psi$ of dimension $2$. To solve for the gauge field, since we have an electric charge at the origin of $AdS_3$. This sources a magnetic field strength of the Chern-Simons field in the angular direction.  We choose 
 a gauge so that only the component $A_\phi$ is non-zero.  We obtain the equations
\begin{eqnarray} \label{electricfield}
\frac{1}{rR} \partial_r A_\phi = \frac{16 \pi G_N }{R}  \frac{\Delta}{R r} \delta ( r) \delta( \phi). 
\end{eqnarray}
Therefore, the solution $A_\phi$ is a constant whose value is fixed by the Stokes theorem
\begin{eqnarray} \label{solelectric}
A_\phi = 8 G_N\frac{ \Delta}{R}.
\end{eqnarray} 

\subsection*{Reading out the expectation value of the boundary operators}

Now that we have the backreacted solution, we obtain the expectation values of the boundary operators. We can read out the stress tensor expectation value in the CFT by writing the metric in the Fefferman-Graham form. We briefly summarise this for convenience. First, we expand the asymptotically $AdS_3$ metric  in (\ref{conical1}) as follows
\begin{eqnarray}
ds^2 = \frac{R^2}{z^2} \left( dz^2 + g_{ij}( z, x)  dx^i dx^j \right),
\end{eqnarray}
then in general  $g_{ij}$  takes the form 
\begin{eqnarray}
g_{ij} ( z, x) = g_{ij}^{(0)} + z^2 g_{ij}^{(2)} + h_{ij} z^2 \log \Big( \frac{z^2}{R^2}  \Big)  + O( z^3). 
\end{eqnarray} 
The expectation of the stress tensor of the boundary CFT is given by 
\begin{eqnarray}
\langle T_{ij} \rangle_{{\rm FG} , AdS_3}= \frac{R}{ 4G_N} g_{ij}^{(2) }.
\end{eqnarray}
Carrying this calculation on the metric in (\ref{conical1}) leads to  the following non-vanishing components for the expectation value of the boundary stress tensor
\begin{eqnarray} \label{e1}
\langle T_{tt} \rangle_{{\rm FG}, AdS_3} &=&  -\frac{R}{ 4 G_N  } \frac{ R^2 - \tilde m}{ 2 R^4}   \\ \nonumber
&=& \frac{\Delta}{R^2} - \frac{c}{12 R^2} \\ \nonumber
&=& \langle T_{\phi \phi} \rangle_{{\rm FG}, AdS_3}.
\end{eqnarray}
To obtain the second line, we have used (\ref{mass-dim}) and the Brown-Henneaux formula (\ref{brownhen}). 
In Fefferman-Graham coordinates, the $U(1)$ gauge field takes the form \cite{Kraus:2006wn}
\begin{eqnarray}
\lim_{z\rightarrow 0 } A_\mu (z, x)  = A_\mu^{(0)}(x) +  z^2 A_\mu^{(2)}(x) + \cdots 
\end{eqnarray}
We evaluate the expectation value of the charge  density and current by using the holographic formula \cite{Kraus:2006wn}
\begin{eqnarray} \label{holcharge}
\langle j^t \rangle_{AdS_3}  = \frac{1}{ 8 G_N } A_t ^{(0)},
\qquad 
\langle j^\phi \rangle _{AdS_3} =  \frac{1}{ 8 G_N   } A_\phi^{(0)}.
\end{eqnarray}
We have removed the factor of $R$, which occurs in the numerator, so that the gauge fields are restored to their canonical dimension of one in the action (\ref{3daction}). Observe that the gauge fields $A_t, A_\phi$  in the action are dimensionless. Reading out the expectation value of the charge density  and current using  the solution  (\ref{solelectric}) and (\ref{holcharge}), we obtain 
\begin{eqnarray} \label{e2}
\langle j^t  \rangle_{\rm AdS_3}  = 0, \qquad \qquad \langle j^\phi\rangle_{\rm AdS_3}  = \frac{\Delta}{R}.
\end{eqnarray}
The expectation value of the scalar operator $\Psi$ of conformal dimension $\Delta_\Psi=2$ is obtained by reading out the fall off behaviour in the dual scalar field $\psi$ in (\ref{solpsi}) 
\begin{eqnarray}
\lim_{r\rightarrow \infty} \psi  (r) = \psi_0  + \psi_1 \frac{ R^2 }{r^2}  + \cdots 
\end{eqnarray}
then   \footnote{It is conventional to define the coordinate $z=R^2/r$ and read out the coefficient of $z^{\Delta_\Psi}$ as the expectation value of the dual operator. 
This explains the additional factor of $1/R^2$  in (\ref{scalexpect}) which results in the correct scaling behaviour for the expectation value. }
\begin{eqnarray} \label{scalexpect}
\langle \Psi \rangle _{\rm AdS_3} =   ( 2\Delta_\Psi  -2)  \frac{ R }{ 4G_N }  \frac{  \psi_1 }{R^2}. 
\end{eqnarray}
Reading out $\psi_1$   from the solution in (\ref{solpsi}) and using  (\ref{scalexpect}) we obtain
\begin{eqnarray} \label{e3}
\langle \Psi\rangle_{\rm AdS_3}  =   - \frac{\Delta}{R^2}.
\end{eqnarray}

\subsubsection*{The CFT dual of the backreacted solution}

From the expectation values of the stress tensor, the $U(1)$ current and the marginal operator $\Psi$ in (\ref{e1}), (\ref{e2}) and (\ref{e3}), obtained from the back reacted solution, it is clear that the CFT is in an excited state with a scalar operator ${\cal O}$ of conformal dimension $\Delta_{\cal O } = \Delta$. To confirm this, we evaluate the following expectation value of the state created by inserting the operator ${\cal O}$ on the cylinder at $t\rightarrow -\infty$.  One approach to do this  would be to evaluate the expectation values in the Euclidean cylinder and use the transformation 
\begin{equation}
z_{\rm plane} =  \exp\Big(-  i\frac{  w}{R} \Big) ,  \qquad \qquad  w = \phi + i \tau, 
\end{equation}
and then transform to the Minkowski theory by taking 
$\tau  = i t$ . Note that the cylinder is of radius $R$.   Performing this, 
we obtain 
\begin{eqnarray}
\frac{ \langle {\cal O } | T_{tt} (0, 0) |{\cal O }    \rangle  }{ \langle {\cal O } |{\cal O } \rangle } &= &
- \frac{ \langle {\cal O } | T_{ww}(0)   + T_{\bar w\bar w} (0 )  |{\cal O }    \rangle  }{ \langle {\cal O } |{\cal O } \rangle } 
\\ \nonumber 
&=& \frac{\Delta}{R^2} - \frac{c}{12 R^2}.
\end{eqnarray}
The shift by the central charge occurs due to the Schwarzian in the transformation of the stress tensor from the plane to the cylinder.  The left and right conformal dimensions of the scalar operator are the same and equal to $\frac{\Delta}{2}$. 
Since the left and right weights are the same, we get 
\begin{eqnarray}
\frac{ \langle {\cal O } | T_{\phi\phi} (0, 0)  |{\cal O }    \rangle  }{ \langle {\cal O } |{\cal O } \rangle }  =
\frac{ \langle {\cal O } | T_{ww} (0) +  T_{\bar w\bar w} (0)  |{\cal O }    \rangle  }{ \langle {\cal O } |{\cal O } \rangle }
&= & \frac{\Delta}{R^2} - \frac{c}{12 R^2}.
\end{eqnarray}
Let the scalar operator be  charged under a $U(1)$ conserved current in the CFT with charges 
$\big( \frac{\Delta}{2},  \frac{\Delta}{2} \big)$ both for the left and right moving $U(1)$ currents, we obtain 
\begin{eqnarray}
\frac{ \langle {\cal O } | j_{t}(0, 0)  |{\cal O }    \rangle  }{ \langle {\cal O } |{\cal O } \rangle } &= &
- i \frac{ \langle {\cal O } | j_{w}  (0) - j_{\bar w}  (0) |{\cal O }    \rangle  }{ \langle {\cal O } |{\cal O } \rangle }  
\\ \nonumber 
&=&0.
\end{eqnarray}
Similarly, we have 
\begin{eqnarray}
\frac{ \langle {\cal O } | j_\phi (0, 0)   |{\cal O }    \rangle  }{ \langle {\cal O } |{\cal O } \rangle } &= & 
 \frac{ \langle {\cal O } | j_{w}  (0) + j_{\bar w}  (0) |{\cal O }    \rangle  }{ \langle {\cal O } |{\cal O } \rangle } \\ \nonumber
 &=& \frac{\Delta}{R} 
\end{eqnarray}
Now the expectation value of the marginal operator $\Psi$ in the state $|{\cal O}\rangle$ is given by 
\begin{eqnarray}
\frac{\langle {\cal O }|  \Psi ( 0, 0 ) | {\cal  O } \rangle } { \langle {\cal O } |{\cal O } \rangle } = 
\frac{ C_{ {\cal O }{\cal O } \Psi } }{R^2}. 
\end{eqnarray}
where $C_{ {\cal O }{\cal O } \Psi }$ is the OPE coefficient of these operators. 
Comparing with the result for the expectation value  from holography in (\ref{e3})
allows us to  identify the $3$ point function as 
\begin{eqnarray} \label{3pointfn}
C_{ {\cal O }{\cal O } \Psi } = -\Delta.
\end{eqnarray}

\subsection{In falling spinning geodesics in $BTZ \times S^3 \times M$ as quenches }
\label{quenchesbtz}

\begin{figure}[h]
    \begin{subfigure}{.46\linewidth}
    \centering
 \includegraphics[width=.8\linewidth]{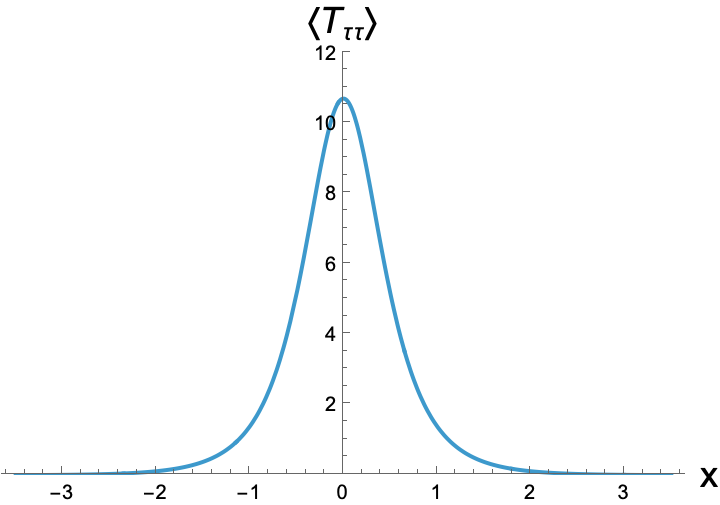}
  \caption{ }
 \label{fig 1 a}
 \end{subfigure}\hfill
 \begin{subfigure}{.46\linewidth}
  \centering
 \includegraphics[width=.8\linewidth]{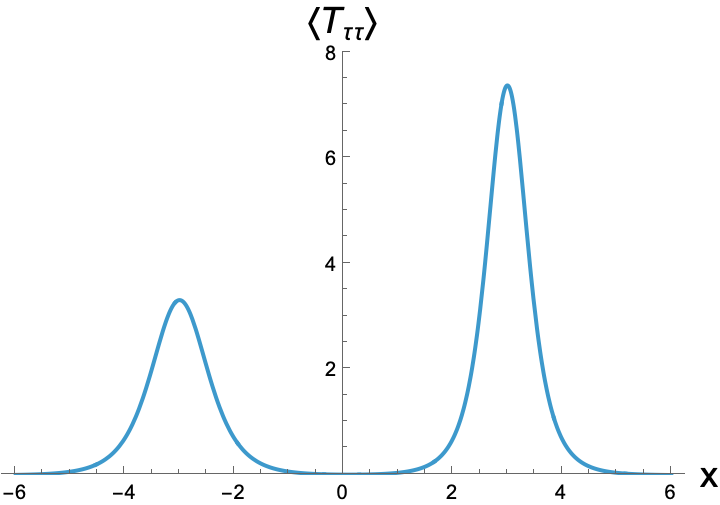}
  \caption{}
\label{fig 1 b}
\end{subfigure}
\caption{Expectation value of the energy density of the local quench.  Panel (a) shows $\langle T_{\tau\tau} \rangle - \frac{c\pi^2}{3\beta^2}$ at $\tau = 0$, while panel (b) shows it  at $\tau = 3$. Both plots are generated for the parameter values $\Delta = 2$, $\eta_1 - \eta_2 = 0.6$, and $\eta_1 + \eta_2 = 1$ with $\frac{2\pi }{\beta} =1$. 
The profile initially consists of a single pulse, which subsequently splits into two pulses propagating at the speed of light. The height
and width of the pulses are controlled by the parameters $\eta_1$, $\eta_2$, and the conformal dimension $\Delta$ of the primary operator. The pulses become symmetric when $\eta_2 =0$. }
\label{AsymmetricQuenchStressTensor}
\end{figure}

The CFT dual to the BTZ geometry is a thermal CFT; therefore, we would expect the geometry with in-falling geodesics or strings into the BTZ to be excitations in the thermal CFT.  We would like to identify these excitations in the thermal CFT. In this section, we first construct the backreacted geometry of the in-falling geodesic in BTZ and spinning on $S^3$ and then use it to find the expectation values of the stress tensor, the $U(1)$ current and the marginal operator in the dual thermal CFT. 
We have shown that the coordinate transformation (\ref{coordinate transformation for boosted in-falling geodesic}) takes the particle at the origin of $AdS_3$ and spinning on $S^3$ to an in-falling geodesic in BTZ and spinning on $S^3$. Therefore, we  use this coordinate transformation on the backreacted solution of the 
particle  spinning particle at the origin of $AdS_3$ given in equations (\ref{conical1}), (\ref{solpsi}), (\ref{conical1}) 
to obtain the backreacted solution of the particle in falling in the BTZ background. It is convenient to work with the following re-definition of the radial coordinate
\begin{eqnarray}
\rho = \frac{1}{z}\ . 
\end{eqnarray}
Now the radial coordinate $z$ has the dimensions of length, since $\rho$ has dimensions of mass. The BTZ metric then becomes  
\begin{eqnarray}
ds^2 = \frac{R^2}{z^2}\lrt{-\lr{1- M^2 z^2}d\tau^2 + \frac{dz^2}{1- M^2z^2} + dx^2}.
\end{eqnarray}
Since we are interested in the background at the leading order in $G_N$, we can proceed by expanding this conical defect metric (\ref{conical1}) to the leading order 
\begin{eqnarray} \label{backplusfluc}
ds^2 & =&  - ( r^2 + R^2) dt^2  + \frac{R^2}{r^2 + R^2} dt^2 + r^2 d\phi^2 + \delta g_{\mu\nu} dx^\mu dx^\nu, \\ \nonumber
 \delta g_{\mu\nu} dx^\mu dx^\nu &=& 
\tilde m \Big( dt^2  + \frac{R^2}{(r^2 + R^2)^2} dr^2 \Big). 
\end{eqnarray}
To make the transformation to the BTZ, we use the relation of the $AdS_3$ coordinates to the embedding  space given in (\ref{coordinate transformation for boosted in-falling geodesic})  and write 
\begin{eqnarray}
dt^2 = \left( \frac{X_0 dX_1 - X_1 dX_0 }{ X_0^2 + X_1^2 } \right)^2, \\ \nonumber
\frac{R^2}{(r^2 + R^2)^2} dr^2  = \frac{ R^2 ( X_2 dX_2 + X_3 dX_3)^2}{ (X_0^2 + X_1^2)^2 (X_2^2 + X_3^2)}. 
\end{eqnarray}
In this form, the perturbative corrections can be easily converted to the BTZ co-ordinates using the relation (\ref{coordinate transformation for boosted in-falling geodesic}). The result is the backreacted metric of an in-falling particle in BTZ. This metric can be expanded near the boundary 
and it of the form
\begin{align}
    ds^2&=\frac{R^2}{z^2}\lrt{-\lr{1- M^2 z^2}d\tau^2 + \frac{dz^2}{1- M^2z^2} + dx^2}  \nonumber\\
    & +\delta g_{zz} (\tau x ) dz^2 +   \delta g_{\tau\tau} ( \tau, x) d\tau^2  +\delta g_{xx}(\tau, x) dx^2
    +\delta g_{\tau x} ( \tau, x ) d\tau dx + \mathcal{O}(z).
\end{align}
Observe that the background $AdS_3$ in (\ref{backplusfluc}) gets converted to  BTZ, while the correction proportional 
to $\tilde m$ will be written as the back reaction to the BTZ background. The expressions for the perturbation are long and cumbersome, but we have kept the leading order at the boundary, whose boundary coordinates are $\tau, x $. We can now go over to Fefferman-Graham coordinates and read out the expectation value of the boundary stress tensor.  This is given by 
\begin{eqnarray}
\langle T_{\tau\tau} \rangle_{\rm FG}  &=&  \frac{R}{4G_N} \Big( \frac{M^2}{2} - \frac{\delta g_{zz}}{2R^2} + 
\frac{\delta g_{\tau\tau}}{R^2} \Big) , \\ \nonumber 
\langle T_{xx} \rangle_{\rm FG}  &=&   \frac{R}{4 G_N} \Big(   \frac{M^2}{2}  + \frac{\delta g_{zz}}{2R^2} 
+ \frac{\delta g_{xx} }{ R^2}  \Big) , \\ \nonumber
\langle T_{\tau x } \rangle_{\rm FG} &=&   \frac{R}{4 G_N } \frac{\delta g_{\tau z }}{2}\ .
\end{eqnarray}
Substituting the perturbations, we obtain 
{\small \begin{eqnarray}
& &\langle T_{\tau\tau} \rangle_{\rm FG} =  \langle T_{xx} \rangle_{\rm FG}  \\ \nonumber
& & = 
\frac{M^2 R}{8G_N }   + \frac{\tilde{m} M^2 }{16G_N R }\Bigg\{
     \frac{1}{\big[\cosh{M(x+\tau)}\cosh{(\eta_1-\eta_2)} - \sinh{(\eta_1-\eta_2)}\big]^2 }  \\ \nonumber
& & \qquad \qquad \qquad\qquad \quad    +
      \frac{1}{\big[\cosh{M(x-\tau)}\cosh{(\eta_1+\eta_2)} - \sinh{(\eta_1+\eta_2)} \big]^2} \Bigg\},
\end{eqnarray} } 
{\small \begin{eqnarray}
\langle T_{\tau x} \rangle_{\rm FG} 
&=&
 \frac{\tilde{m} M^2 }{32G_{\text{\tiny N}} R }\Bigg\{
     \frac{1}{\big[\cosh{M(x+\tau)}\cosh{(\eta_1-\eta_2)} - \sinh{(\eta_1-\eta_2)}\big]^2 }  \\ \nonumber
& & \qquad \qquad \qquad\qquad \quad    -
      \frac{1}{\big[\cosh{M(x-\tau)}\cosh{(\eta_1+\eta_2)} - \sinh{(\eta_1+\eta_2)} \big]^2} \Bigg\}.
\end{eqnarray} }
The expectation value of the stress tensor is conserved and traceless. 
The mass of BTZ is related to its temperature by 
\begin{eqnarray}
M = \frac{2\pi }{\beta}.
\end{eqnarray}
We can use this relation and  the Brown-Henneaux formula in (\ref{brownhen}) to write these expectation values as
{\small \begin{eqnarray} \label{btzstress1}
& &\langle T_{\tau\tau} \rangle_{\rm FG} =  \langle T_{xx} \rangle_{\rm FG}  \\ \nonumber
& & = \frac{c}{3} \frac{\pi^2}{\beta^2}  + \frac{\Delta}{2} \Big(  \frac{2\pi}{\beta} \Big)^2  \Bigg\{
\frac{1}{\Big[\cosh{\frac{2\pi}{\beta} (x+\tau)}\cosh{(\eta_1-\eta_2)} - \sinh{(\eta_1-\eta_2)}\Big]^2}   \\ \nonumber
&& \qquad\qquad\quad\qquad\qquad\quad
+ \frac{1}{\Big[ \cosh{\frac{2\pi}{\beta} (x-\tau)}\cosh{(\eta_1+\eta_2)} - \sinh{(\eta_1+\eta_2)} \Big]^2} \Bigg\},
\end{eqnarray}
\begin{eqnarray} \label{btzstress2}
\langle T_{\tau x} \rangle_{\rm FG} 
&=&
\frac{\Delta}{4} \Big(  \frac{2\pi}{\beta} \Big)^2  \Bigg\{
\frac{1}{\Big[\cosh{\frac{2\pi}{\beta} (x+\tau)}\cosh{(\eta_1-\eta_2)} - \sinh{(\eta_1-\eta_2)}\Big]^2}   \\ \nonumber
&& \qquad\qquad\quad\qquad\qquad\quad
- \frac{1}{\Big[ \cosh{\frac{2\pi}{\beta} (x-\tau)}\cosh{(\eta_1+\eta_2)} - \sinh{(\eta_1+\eta_2)} \Big]^2} \Bigg\}.
\end{eqnarray} }
In Figure \ref{AsymmetricQuenchStressTensor}, we have plotted the expectation value of the energy density of the
quench at instances of time.
The pulse splits asymmetrically and quenches travel at the speed of light. 
For $\eta_1>>1, \eta_2\sim 1$, the height of the energy densities of the pulses is approximately given by 
\begin{eqnarray} \label{h1h2}
{\rm Height\;of\;pulse\;1} \sim  \frac{\Delta}{2} \big( \frac{2\pi }{\beta} \big)^2 e^{ 2 ( \eta_1 - \eta_2)}  , \\ \nonumber
{\rm Height\;of\;pulse\;2} \sim  \frac{\Delta}{2} \big( \frac{2\pi }{\beta} \big)^2 e^{ 2 ( \eta_1 + \eta_2) }. 
\end{eqnarray}
To get an idea of the width of the pulses, it is convenient to take the limit $\beta>> x, \tau$ and $\eta_1 >>1, \eta_2 \sim 1$, then the widths of the two pulses are given by 
\begin{eqnarray} \label{widths} \label{w1w2}
{\rm Width \;of\;pulse\;1} \sim   \frac{\beta}{2\pi } e^{- ( \eta_1 - \eta_2)} , \\ \nonumber
{\rm Width \;of\;pulse\;2} \sim   \frac{\beta}{2\pi } e^{- ( \eta_1 + \eta_2)}.
\end{eqnarray}
Therefore, the taller pulse is narrower than the smaller one, which can be clearly seen in the figure \ref{AsymmetricQuenchStressTensor}. 
Ignoring the thermal energy of the vaccum, 
the  total energy carried by the left moving pulse is given by 
\begin{eqnarray}
\hbox{Energy of pulse 1} &=&
\frac{\Delta}{2} \Big(  \frac{2\pi }{\beta} \Big)^2  \int_{-\infty}^\infty  dx \Bigg\{
\frac{1}{\Big[\cosh{\frac{2\pi}{\beta} (x+\tau)}\cosh{(\eta_1-\eta_2)} - \sinh{(\eta_1-\eta_2)}\Big]^2}  \Bigg\},  \nonumber \\
&=& \frac{2\pi \Delta }{\beta}  \Big[ 1+ 2 \arctan( e^{\eta_1-\eta_2} ) \sinh(\eta_1 - \eta_2)  \Big]
\end{eqnarray}
Again in the limit $\eta_1>>1, \eta_2\sim 1$ we see the left move pulse has the energy
\begin{eqnarray} \label{e1}
\hbox{Energy of pulse 1}|_{\eta_1>>1, \eta_2\sim 1}  &=& 
\frac{\pi^2  \Delta}{\beta} e^{ (\eta_1 - \eta_2) } .
\end{eqnarray}
Similarly, the right moving pulse has the energy 
\begin{eqnarray} \label{e2}
 \hbox{Energy of pulse 2}|_{\eta_1>>1, \eta_2\sim 1}  &=& 
\frac{\pi^2  \Delta}{\beta} e^{ (\eta_1 + \eta_2) } .
\end{eqnarray}
From equations ( \ref{h1h2}), ( \ref{w1w2}), ( \ref{e1}),  (\ref{e2})
we  see that the parameters $(\eta_1- \eta_2)$,  $(\eta_1 + \eta_2)$ determine parameters of the left and right moving 
pulses respectively.

We now evaluate the expectation value of the $U(1)$  current in the backreacted geometry of the in-falling geometry. For this, it is convenient to use the coordinate transformation of the background gauge field found for the particle in $AdS_3$ to BTZ  from (\ref{coordinate transformation for boosted in-falling geodesic}) directly. 
This transformation is given by 
{\small \begin{eqnarray} \label{ads3btz}
t &=& \arctan \left[
\frac{\sqrt{\rho^2 -M^2} \sinh( M \tau) \cosh \eta_2 -  \rho \sinh\eta_2 \sinh(  M x ) }
{\rho \cosh\eta_1 \cosh( M x) -  \sqrt{\rho^2 - M^2} \cosh ( M \tau) \sinh \eta_1 }
\right], \\ \nonumber
\phi &=& \arctan\left[ \frac{ \rho \sinh( M x) \cosh \eta_2 - \sqrt{ \rho^2 - M^2} \sinh(M \tau)  \sinh\eta_2}
{\sqrt{\rho^2 - M^2} \cosh ( M \tau) \cosh\eta_1 -  \rho \sinh\eta_1 \cosh( M x) } \right], \\ \nonumber
r &=& \frac{R}{M} \Big[ 
 \big( \rho \sinh( M x) \cosh \eta_2 - \sqrt{ \rho^2 - M^2} \sinh(M \tau)  \sinh\eta_2 \big)^2  \\ \nonumber
& & \quad\qquad  + \big( \sqrt{\rho^2 -M^2} \sinh( M \tau) \cosh \eta_2 -  \rho \sinh\eta_2 \sinh(  M x ) \big)^2 
 \Big]^{\frac{1}{2} }. 
\end{eqnarray} }
Using these transformation we can use the field strength of the backreacted solution in $AdS_3$ in (\ref{solelectric}) to obtain the expectation value charge density at the boundary of the BTZ background. This results in 
{\small 
\begin{eqnarray} \label{btzcurrent1}
&&\langle j_\tau \rangle_{\rm BTZ} =  \lim_{\rho\rightarrow \infty} \frac{R}{8 G_N } A_\tau  = 
\lim_{\rho\rightarrow \infty}  \frac{R}{8 G_N } \frac{\partial \phi}{\partial \tau} A_\phi   \\ \nonumber
&=& 
 \frac{\Delta M}{2 }
  \ltb{\frac{1}{ \cosh{ M(x+\tau)}\cosh{(\eta_{1} - \eta_{2})} - \sinh(\eta_{1} - \eta_{2})} - \frac{1}{ \cosh{ M(x-\tau)}\cosh{(\eta_{1} + \eta_{2})} - \sinh(\eta_{1} + \eta_{2})}} \\ \nonumber
&=&  \frac{\Delta }{2 } \frac{2\pi }{\beta} 
  \ltb{\frac{1}{ \cosh{ \frac{2\pi}{\beta} (x+\tau)}\cosh{(\eta_{1} - \eta_{2})} - \sinh(\eta_{1} - \eta_{2})} - \frac{1}{ \cosh{ \frac{2\pi }{\beta } (x-\tau)}\cosh{(\eta_{1} + \eta_{2})} - \sinh(\eta_{1} + \eta_{2})}}.
 \end{eqnarray}}
 In the second line, we have written the expression in terms of the temperature of the BTZ background. In BTZ coordinates, the gauge field had dimensions of inverse length; therefore, there is an additional factor of $R$ along with $G_N$ as compared to  (\ref{holcharge}), which makes the overall factor the level of the Chern-Simons field as expected. \footnote{Recall the coordinate $t$ in $AdS_3$ is dimensionless while the coordinate $\tau$ in  BTZ has the dimensions of length. }. 
 Similarly, the spatial component of the current is given by 
{\small  \begin{eqnarray} \label{btzcurrent2}
& &  \langle j_{x} \rangle_{\rm BTZ} = 
\lim_{\rho\rightarrow \infty} \frac{R}{8 G_N } A_x  = 
\lim_{\rho\rightarrow \infty}  \frac{R}{8 G_N } \frac{\partial \phi}{\partial x} A_\phi
\ \\ \nonumber
&=& \frac{\Delta M}{2 }
  \ltb{\frac{1}{ \cosh{ M(x+\tau)}\cosh{(\eta_{1} - \eta_{2})} - \sinh(\eta_{1} - \eta_{2})} + \frac{1}{ \cosh{ M(x-\tau)}\cosh{(\eta_{1} + \eta_{2})} - \sinh(\eta_{1} + \eta_{2})}} \\ \nonumber
&=&  \frac{\Delta }{2 } \frac{2\pi }{\beta} 
  \ltb{\frac{1}{ \cosh{ \frac{2\pi}{\beta} (x+\tau)}\cosh{(\eta_{1} - \eta_{2})} - \sinh(\eta_{1} - \eta_{2})} + \frac{1}{ \cosh{ \frac{2\pi }{\beta } (x-\tau)}\cosh{(\eta_{1} + \eta_{2})} - \sinh(\eta_{1} + \eta_{2})}}.
\end{eqnarray} }
 
 \begin{figure}[h]
    \begin{subfigure}{.46\linewidth}
     \centering
 \includegraphics[width=.8\linewidth]{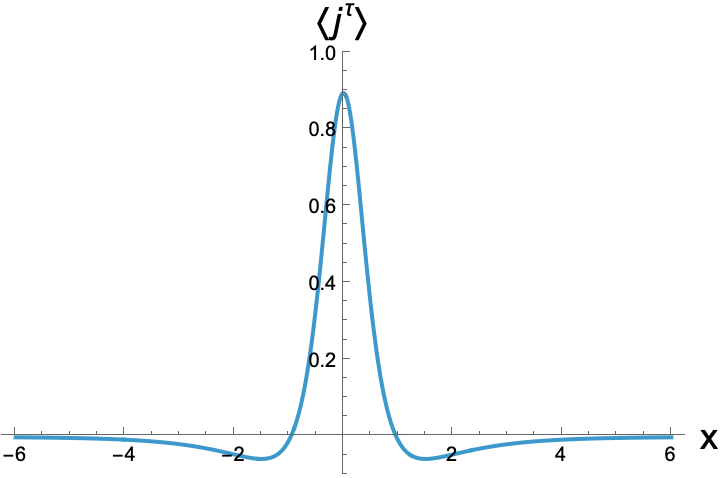}
 \caption{}
 \label{fig 2 a}
 \end{subfigure}\hfill
 \begin{subfigure}{.46\linewidth}
  \centering
 \includegraphics[width=.8\linewidth]{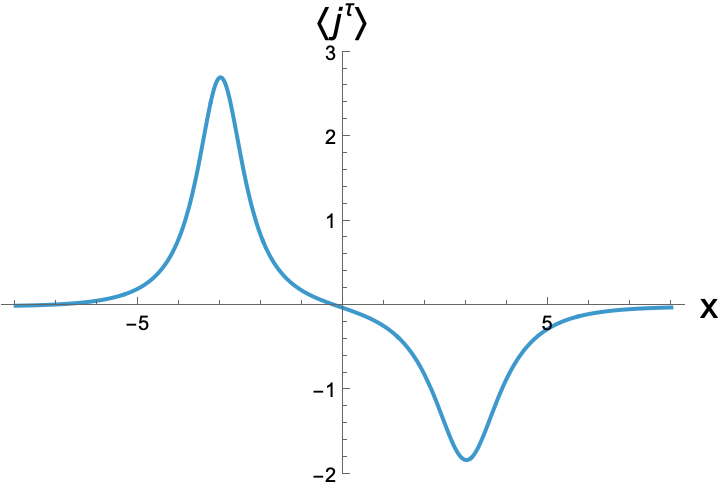}
 \caption{}
\label{fig 2 b}
\end{subfigure}
\caption{Profile of the expectation value of the charge at different times. Panel (a) shows $\langle j^\tau \rangle$ at $\tau = 0$, while panel (b) shows $\langle j^\tau \rangle$ at $\tau = 3$. Both plots are generated for $\Delta = 2$, $\eta_1 - \eta_2 = 1$, and $\eta_1 + \eta_2 = 0.6$ with $\frac{2\pi }{\beta} =1$.  Here, we have interchanged the value of the parameters so that the pulse 
at $\tau =0$ is mostly positive. 
The qualitative behaviour of the profile remains the same as in the case of the energy density. The height and width of the pulses are determined by the parameters $\eta_1$, $\eta_2$, and the conformal dimension $\Delta$ of the primary operator.} \label{chargefig}
\end{figure}

\begin{figure}[h]
    \begin{subfigure}{.46\linewidth}
     \centering
 \includegraphics[width=.8\linewidth]{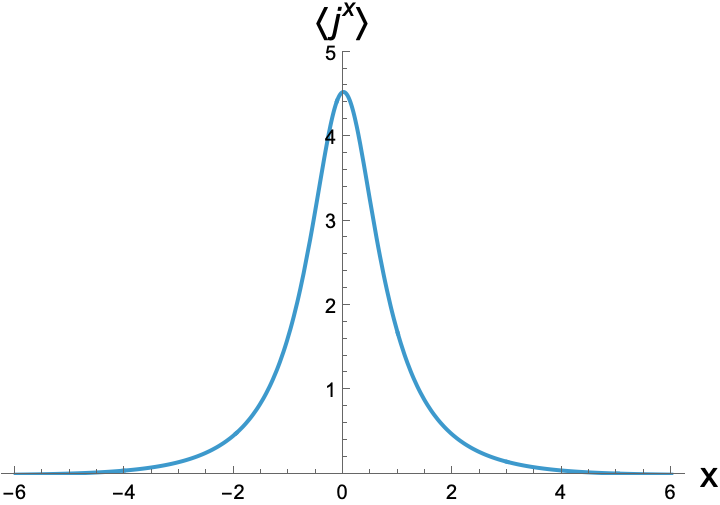}
 \caption{}
 \label{fig 3 a}
 \end{subfigure}\hfill
 \begin{subfigure}{.46\linewidth}
  \centering
 \includegraphics[width=.8\linewidth]{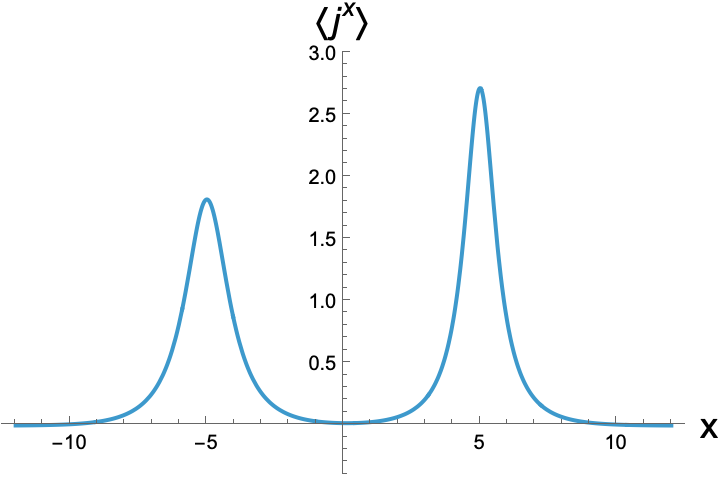}
 \caption{}
\label{fig 3 b}
\end{subfigure}
    \caption{Profile of the expectation value of the current at different times. Panel (a) shows $\langle j^x \rangle$ at $\tau = 0$, while panel (b) shows $\langle j^x \rangle$ at $\tau = 5$ with $\frac{2\pi }{\beta} =1$. Both plots are generated for $\Delta = 2$, $\eta_1 - \eta_2 = 0.6$, and $\eta_1 + \eta_2 = 1$. The qualitative behaviour of the profile remains the same as in the previous cases.}
    \label{fig current}
\end{figure}
 
 In figures \ref{chargefig}, \ref{fig current}, we have plotted the profile of the charge density and the current at two instances of time. From $\tau=0$, the pulse splits into two and travels at the speed of light. 
 The quenches are asymmetric with their widths determined as given in equation (\ref{widths}). 
 The heights of both the charge density and current quenches are given by 
 \begin{eqnarray}
&& |{\rm Height\;of\;pulse\;1}| \sim  \frac{\Delta}{2} \big( \frac{2\pi }{\beta} \big) e^{  ( \eta_1 - \eta_2)}  , \qquad 
|{\rm Height\;of\;pulse\;2}|  \sim  \frac{\Delta}{2} \big( \frac{2\pi }{\beta} \big) e^{  ( \eta_1 + \eta_2) } ,  \nonumber \\
&& {\rm for} \;  \eta_1>>1, \eta_2 \sim 1.
 \end{eqnarray} 
 It is also interesting to evaluate the charge carried by  the pulses
 The left moving pulse has charge 
 \begin{eqnarray}
\hbox{Charge of pulse 1} &=&  \frac{\Delta}{2} \frac{2\pi }{\beta} \int_{-\infty}^\infty 
\frac{dx}{ \cosh{ \frac{2\pi}{\beta} (x+\tau)}\cosh{(\eta_{1} - \eta_{2})} - \sinh(\eta_{1} - \eta_{2})} , \nonumber \\
& & = 2\Delta\arctan( e^{\eta_1 - \eta_2} ) . 
 \end{eqnarray}
 In the limit $\eta_1>>1, \eta_2 \sim 1$, the charges carried by the pulses are given by 
 \begin{eqnarray} \label{cha12}
 \hbox{Charge of pulse 1} = 2\Delta\lrt{ \frac{\pi}{2} - e^{- (\eta_1 - \eta_2) }  }, \\ \nonumber
  \hbox{Charge of pulse 2} =2 \Delta\lrt{ \frac{\pi}{2} - e^{- (\eta_1 + \eta_2) }}.
 \end{eqnarray}
 As expected  the total left and right moving charge pulses are characterised by $(\eta_1-\eta_2)$,
 $(\eta_1+ \eta_2 )$ and they are almost equal in the limit of interest.

\begin{figure}[h]
    \begin{subfigure}{.46\linewidth}
    \centering
 \includegraphics[width=.8\linewidth]{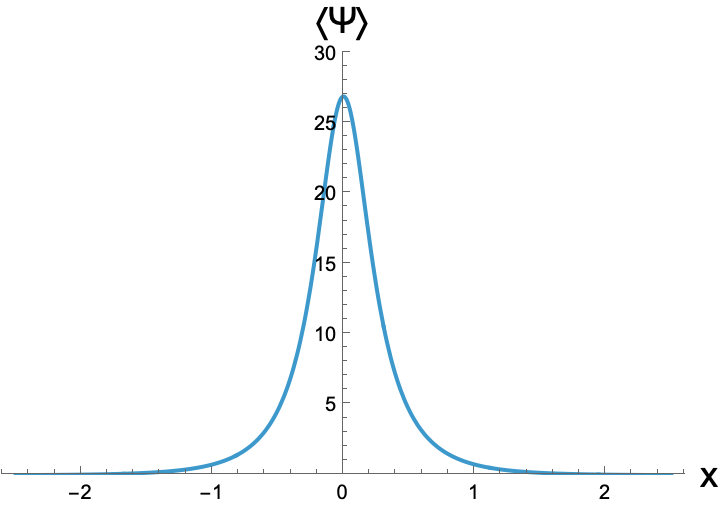}
 \caption{ }
 \label{fig 1 a1}
 \end{subfigure}\hfill
 \begin{subfigure}{.46\linewidth}
  \centering
 \includegraphics[width=.8\linewidth]{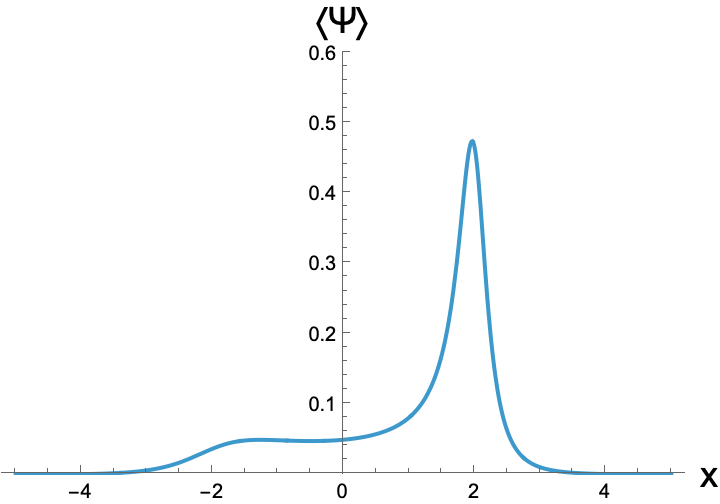}
 \caption{}
\label{fig 1 b1}
\end{subfigure}
\caption{Expectation value of the negative of scalar operator $\Psi$. Panel (a) shows $\langle \Psi \rangle$ at $\tau = 0$, while panel (b) shows $\langle \Psi \rangle$ at $\tau = 2$. Both plots are generated for $\Delta = 2$ , $\eta_1 + \eta_2 = 2$, and  $\eta_1 - \eta_2 = 0.6$ . Note the qualitative contrast between these profiles and those shown in the previous figures. The profile initially consists of a single pulse at all times, though its amplitude decreases and the peak shifts either to the left or to the right depending on the parameter values. For the parameters chosen here, the peak shifts to the right.}
\label{scalarfig}
\end{figure}

The profile of the scalar $\psi$ in the backreacted geometry of the in-falling particle is found by using the coordinate transformation in (\ref{ads3btz}) on the solution of the scalar given in (\ref{solpsi}), which results in 
\begin{eqnarray}
\lim_{\rightarrow \infty} \psi (\rho, \tau, x)  = \psi_0 + \frac{1}{\rho^2} \psi_1(\tau, x) + \cdots 
\end{eqnarray}
The expectation value of the marginal operator $\Psi$ dual to the minimally coupled scalar  $\psi$ sourced by the in-falling particle is given by 
\begin{eqnarray}
\langle \Psi \rangle_{\rm BTZ} = (2\Delta_\Psi - 2) \frac{R}{4 G_N}   \psi_1( \tau, x).
\end{eqnarray}
Note that $\psi_1$ has the dimensions of inverse length squared. Substituting the change of co-ordinates in 
(\ref{ads3btz}) and extracting the coefficient $\psi_1(\tau, x)$, we obtain 
\begin{eqnarray} \label{btzscalar}
&& \langle \Psi \rangle_{\rm BTZ}    \\
&=& \frac{- \Delta M^2}{\lrs{ \cosh{M(x+\tau)}\cosh{(\eta_1-\eta_2)} - \sinh{(\eta_1-\eta_2)}}\lrs{ \cosh{M(x-\tau)}\cosh{(\eta_1+\eta_2)} - \sinh{(\eta_1+\eta_2)}}} \nonumber \\ 
&=& \frac{- \Delta \big( \frac{2\pi}{\beta} \big)^2}{\lrs{ \cosh{\frac{2\pi}{\beta}(x+\tau)}\cosh{(\eta_1-\eta_2)} - \sinh{(\eta_1-\eta_2)}}\lrs{ \cosh{\frac{2\pi}{\beta} (x-\tau)}\cosh{(\eta_1+\eta_2)} - \sinh{(\eta_1+\eta_2)}}}.
\nonumber
\end{eqnarray}
In figure \ref{scalarfig}, we have plotted the expectation value of the scalar operator $|\langle \Psi \rangle |$ at 2 instances of time. 
The expectation value at initial time $\tau =0$ is peaked at the origin, and then the peak moves away from the origin at the speed of light. 
The peak at time $\tau>>\beta$ is approximately given by the expression
\begin{eqnarray}
|\langle \Psi \rangle|_{\rm maximum }  \sim \Delta\big( \frac{2\pi }{\beta} \big)^2 e^{( - \frac{4\pi }{\beta}  \tau + \eta_1 + \eta_2 )}. 
\end{eqnarray}
Thus, the height of the quench decreases in time. The width of the pulse for $\beta>>\tau, x; \eta_1>>1; \eta_2 \sim 1$ is given by 
\begin{eqnarray}
{\rm Width\; of \; scalar\; expectation }  \sim  \frac{\beta}{2\pi } e^{- ( \eta_1 + \eta_2)}.
\end{eqnarray}

\subsubsection*{The density matrix of the in-falling particle in the thermal  CFT}

The CFT dual to the BTZ black hole is a thermal CFT. Therefore, the dual description of the in-falling particle should be an excitation on the thermal density matrix of the CFT. We consider the following state described by a density matrix  for the dual description of the in-falling geodesic in the BTZ background
\begin{eqnarray} \label{densitymat}
\rho =  {\cal N} e^{-i H\tau } {\cal O }(w_1, \bar w_1) \rho_\beta {\cal O }^\dagger ( w_2, \bar w_2) e^{i H \tau}, 
\end{eqnarray}
where $\rho_\beta = e^{-\beta H}$ is the thermal density matrix and the normalisation ${\cal N}$ is chosen so that 
${\rm Tr \rho} =1$. The position coordinates of the operator are given by 
\begin{eqnarray}
  w_1= -i\epsilon_1,\quad w_2= i\epsilon_1,\quad \bar{w}_1= i\epsilon_2,\quad \bar{w}_2= -i\epsilon_2
\end{eqnarray}
with   $\epsilon_i$ real. Such excitations are the generalisations of local quenches considered in \cite{Caputa:2014eta,David:2016pzn} 
for which the parameters $\epsilon_i$ were restricted to  $\epsilon_1=\epsilon_2$.  These parameters determine the width of the local quenches. The coordinates of the operators are labelled in  Lorentzian signature on the cylinder are 
\begin{eqnarray}
(w, \bar w) \equiv( x - \tau,  x+ \tau). 
\end{eqnarray}
The Lorentzian co-ordinates can naturally be continued to Euclidean when $\tau \rightarrow - i \tau_E$. We take the operator ${\cal O}$ to have conformal dimensions $(\frac{\Delta}{2}, \frac{\Delta}{2} )$, with $U(1)$ charges $\big(\frac{\Delta}{2} ,  \frac{\Delta}{2} \big)$. It also has a non-zero $3$ point function with the marginal operator $\Psi$ given by (\ref{3pointfn}). We proceed to evaluate the expectation value of the energy density  in the density matrix (\ref{densitymat}), 
\begin{eqnarray}
\langle T_{\tau\tau}\rangle_{CFT} &=& {\rm Tr}\big[ \rho_\beta T_{\tau\tau} (x- \tau, x + \tau  )  \big] , \\ \nonumber
&=& 
\frac{ \langle {\cal O}^\dagger ( w_2, \bar w_2) T_{\tau\tau} ( x - \tau, x+ \tau)  {\cal O } ( w_1, \bar w_1) \rangle_{\beta} }
{ \langle {\cal O}^\dagger ( w_2, \bar w_2) {\cal O } ( w_1, \bar w_1) \rangle_{\beta} }.
\end{eqnarray}
We use the map from the thermal cylinder to the plane, which is given by 
\begin{eqnarray}\label{thermmap}
z = \exp\big( \frac{ 2\pi }{\beta} w \big), \qquad w = x + i \tau_E, \quad \tau_E = i \tau
\end{eqnarray}
to obtain thermal expectation values. 
This results in 
\begin{eqnarray}
\langle T_{\tau\tau} \rangle_{CFT} &=&  \frac{\pi^2c}{3\beta^2}  -\frac{4\pi^2}{\beta^2}\lrt{ z^2 
\frac{\langle \mathcal{O}^\dagger(z_2, \bar z_2 ) T(z)\ \mathcal{O}(z_1, \bar z_1)\rangle_\beta} {
\langle\mathcal{O}^\dagger (z_2, \bar z_2 )\ \mathcal{O}(z_1, \bar z_1 )\rangle_\beta}+ \bar{z}^2\frac{\langle \mathcal{O}^\dagger ( z_2, \bar{z}_2  ) \bar{T}(\bar{z})\ \mathcal{O}(z_1, \bar z_1 )\rangle_\beta}{\langle\mathcal{O}^\dagger(z_2, \bar z_2 )\ \mathcal{O}(z_1, \bar{z}_1 )\rangle_\beta}}.  \nonumber \\
\end{eqnarray}
To obtain this expression, we have used the transformation of the cylinder to the plane, which results in the Schwarzian contribution along with the three-point function of the stress tensor with the two primaries. The three-point function is determined by the stress tensor Ward identity. Here we need to substitute the coordinates on the plane in terms of the cylinder using the map (\ref{thermmap}). This results in 
\begin{eqnarray}
\langle T_{\tau\tau} \rangle_{CFT} &=& 
\frac{\pi^2 c}{3\beta^2}+ \frac{4\pi^2\Delta }{2\beta^2} \lrt{\frac{\sin^2{\frac{2\pi \epsilon_1}{\beta}}}{\lrs{\cosh{\frac{2\pi}{\beta}(x +\tau)} - \cos{\frac{2\pi \epsilon_1}{\beta}}}^2} +\frac{\sin^2{\frac{2\pi \epsilon_2}{\beta}}}{\lrs{\cosh{\frac{2\pi}{\beta}(x -\tau)} - \cos{\frac{2\pi \epsilon_2}{\beta}}}^2}} \nonumber.
\\
\end{eqnarray}
On comparing the boundary stress expectation value obtained from the backreacted geometry of the in-falling particle  in (\ref{btzstress1}) and the above expectation value of the stress tensor in the state (\ref{densitymat}), we see that upon making the identifications 
\begin{eqnarray}\label{gravcftiden}
\tanh{(\eta_1-\eta_2)}= \cos{\frac{2\pi\epsilon_1}{\beta}}, 
    \qquad \tanh{(\eta_2+\eta_1)}= \cos{\frac{2\pi\epsilon_2}{\beta}},
\end{eqnarray} they precisely agree. 
It is useful to examine this identifications in the limit $\eta_1>>1, \eta_2 \sim 1$ for which we obtain 
\begin{eqnarray} \label{limitrelation}
e^{-(\eta_1 - \eta_2)} \sim  \frac{\pi\epsilon_1}{\beta}, \qquad\qquad e^{-(\eta_1 + \eta_2) } \sim  \frac{\pi\epsilon_2}{\beta}
\end{eqnarray} 
Observe that in this limit, the energies of the pulses in (\ref{e1}), (\ref{e2}) are regularized by these widths 
\begin{eqnarray}
\hbox{Energy of pulse 1}|_{\epsilon_1 \rightarrow 0}  = \frac{\pi \Delta}{\epsilon_1}, \qquad
\hbox{Energy of pulse 2}|_{\epsilon_2 \rightarrow 0}  = \frac{\pi \Delta}{\epsilon_2}
\end{eqnarray}
Note that the width of the two  pulses from (\ref{w1w2}) are $\epsilon_1$ and $\epsilon_2$ respectively 
and therefore we observe that  energy is distributed so that the product of total energy of the pulse 
times the width of the pulse is constant and proportional to the dimension of the operator.

We can repeat the same calculation in the CFT for the expectation value $\langle T_{\tau x} \rangle_{CFT} $, and it 
too agrees with (\ref{btzstress2}) upon the identifications (\ref{gravcftiden}). Let us proceed to the expectation value of the charge density. We obtain 
\begin{eqnarray}
\langle j_\tau \rangle_{\rm CFT} &=& {\rm Tr} \big[ \rho_\beta j_\tau  ( x- \tau, x+ \tau)  \big] , \\ \nonumber
&=& - i 
\lrt{\frac{\langle {\mathcal{O}^\dagger (w_2, \bar w_2 )j (w)\mathcal{O}(w_1, \bar w_1 )}\rangle_\beta }{\langle{\mathcal{O}^\dagger (w_2, \bar w_2)\mathcal{O}(w_1, \bar w_1 )}\rangle_\beta }-
\frac{\langle {{\mathcal{O}^\dagger }( w_2 , \bar{w}_2)\bar{j}(\bar{w}){\mathcal{O}}(w_1, \bar{w}_1)}\rangle_\beta}{\langle {{\mathcal{O}^\dagger }(w_2, \bar{w}_2)\bar{\mathcal{O}}(w_1, \bar{w}_1)}\rangle_\beta}}.
\end{eqnarray}
We can evaluate the first three-point function using the fact that the holomorphic current is a dimension $(1, 0 )$ operator and the operator ${\cal O }$ has the holomorphic charge $\frac{\Delta}{2}$ and using the cylinder to plane map. A similar approach can be done for the anti-holomorphic three-point function. This results in the following expectation value
\begin{eqnarray}
\langle j_\tau \rangle_{{ \rm CFT} \beta } = \frac{\Delta}{2} \frac{2\pi }{\beta} 
\lrt{ \frac{\sin{\frac{2\pi}{\beta}\epsilon_1}}{\cosh{\frac{2\pi}{\beta}(x+\tau)}-\cos{\frac{2\pi}{\beta}\epsilon_1}} -\frac{\sin{\frac{2\pi}{\beta}\epsilon_2}}{\cosh{\frac{2\pi}{\beta}(x-\tau)}-\cos{\frac{2\pi}{\beta}\epsilon_2}} }.
\end{eqnarray}
Again, we see that this precisely coincides with the expectation value of the $U(1)$ current obtained from the backreacted solution of the in-falling geodesic in  
(\ref{btzcurrent1}) using the identification (\ref{gravcftiden}). A similar calculation can also be done for the expectation value of the $j_x$ component, resulting in 
the holographic result (\ref{btzcurrent2}).  From (\ref{cha12}) and the relation (\ref{limitrelation}), 
we see that in the small 
$\epsilon_1, \epsilon_2$ limit, the total charge carried by the left and right pulses are almost equal.

Finally, let us evaluate the expectation value of the marginal operator $\Psi$ of dimension $(1, 1)$ in the state 
(\ref{densitymat}). 
\begin{eqnarray}
\langle \Psi \rangle_{\rm CFT} &=& {\rm Tr} \big(  \rho_\beta \Psi ( x - \tau, x + \tau) \big ) , \\ \nonumber
&=& \frac{ \langle  { \cal O} ^\dagger ( w_ 2, \bar w_2) \psi ( x - \tau, x+ \tau) {\cal O } ( w_1, \bar w_1) 
\rangle_\beta  }
{\langle {{\mathcal{O}^\dagger }(w_2, \bar{w}_2)\bar{\mathcal{O}}(w_1, \bar{w}_1)}\rangle_\beta}.
\end{eqnarray}
Evaluating the correlator, by conformally mapping the cylinder to the plane, we obtain 
\begin{eqnarray}
\langle \Psi \rangle_{\rm CFT}  =  -\Delta
\big( \frac{2\pi}{\beta} \big)^2 
\lrt{ \frac{\sin{\frac{2\pi}{\beta}\epsilon_1}\sin{\frac{2\pi}{\beta}\epsilon_2}}{\lrs{\cosh{\frac{2\pi}{\beta}(x+\tau)}-\cos{\frac{2\pi}{\beta}\epsilon_1}}\lrs{\cosh{\frac{2\pi}{\beta}(x-\tau)} - \cos{\frac{2\pi}{\beta}\epsilon_2}}}}.
\end{eqnarray}
Comparing with (\ref{btzscalar}), we see that this expression for the expectation value of the marginal operator $\Psi$ precisely agrees with that evaluated from the backreacted geometry of an in-falling geodesic upon using the relations in 
(\ref{gravcftiden}). In conclusion, we have shown that the spinning in falling particle in the BTZ geometry is dual to a local quench carrying $U(1)$ R-charge corresponding to the angular momentum in the $S^3$. In general, the left and right moving pulses of the 
quench are not symmetric. The quench also carried the expectation value of a marginal operator.

\subsection{ Quenches dual to the   circular string}

Let us consider the circular string solution in $AdS_3\times S^3 \times M $ given by 
\begin{eqnarray}
&&t(\sigma^0, \sigma^1) = \kappa \sigma^0,  \quad  r( \sigma^0, \sigma^1) = 0 , \quad
\phi( \sigma^0, \sigma^1) = 0 , \\ \nonumber
&&\psi =  \tan^{-1}\lr{\sqrt{\frac{W-qm}{W + qm}}} \equiv \psi_0 , \quad 
\phi_1=\sigma^0\lr{W + qm} + m\sigma^1,\quad
    \phi_2=\sigma^0\lr{W - qm}- m\sigma^1,
    \\ \nonumber
   && X(\sigma^0, \sigma^1) |_M = x_m = {\rm constant}. 
\end{eqnarray}
with $m \in \mathbb{Z}$. Here, we have used the parametrisation 
(\ref{parameterization of su2 group element}), for the metric of $S^3$ given in (\ref{mets3}). 
\begin{eqnarray}
\kappa^2 = \omega^2 + m^2,  \qquad W^2 = w^2 + q^2 m^2.
\end{eqnarray}
The angular momentum and the dispersion relation are given  by 
\begin{eqnarray}
J_1= J_2 = \frac{k}{2} ( W - qm ), \qquad J \equiv J_1 + J_2, \\ \nonumber
\Delta = k \kappa  =\sqrt{ ( J + q km)^2 + k^2 m^2 ( 1- q^2) }.
\end{eqnarray}
The world sheet couples to the target space metric by the action given in (\ref{swscoupling}). 
We substitute the solution and derive the spacetime stress tensor. To un-clutter the equations, we first  define the delta function 
\begin{eqnarray}
\hat \delta  \equiv  \frac{ \delta(r)\delta( \phi) }{r}
\frac{  \delta(\psi -\psi_0)  \delta ( \phi_1 + \phi_2 - \frac{ 2 W }{\kappa } t) }{\sqrt{g_{S^3} }}
\frac{  \delta^4( X_M - x_M) }{\sqrt{G_M}}.
\end{eqnarray}
Then the stress tensor sourced by the circular string solution for $m \neq 0$ is given by 
\begin{eqnarray} \label{circular10d}
&& T^{t\,t} =  \frac{\Delta}{2\pi m R}  \hat \delta, 
\qquad  T^{t\, \phi_1} = \frac{ J + 2 k qm}{ 2\pi m R}  \hat \delta, \\ \nonumber
&& T^{t\, \phi_2}  = \frac{J}{2\pi m R}  \hat \delta, 
 \qquad 
T^{\phi_1\phi_1} = \frac{ ( J +  2 k qm )^2 -k^2m^2 }{ 2\pi m R \Delta } \hat \delta, \\ \nonumber
&& T^{\phi_2\phi_2} =   \frac{J^2 -k^2 m^2 }{ 2\pi m R\Delta}   \hat \delta, \qquad
T^{\phi_1\phi_2} = \frac{J  (J +  2 k qm ) -k^2 m^2 } { 2\pi m R\Delta}  \hat \delta.
\end{eqnarray}
We have to treat the case $m =0$ for which there is no world sheet spatial dependence, separately. For this case, the delta function would become 
\begin{eqnarray}
\hat \delta \rightarrow \check\delta \equiv 
\frac{ \delta(r)\delta( \phi) }{r}
\frac{  \delta(\psi -\psi_0 )  \delta ( \phi_1 - \frac{W}{\kappa } t) \delta ( \phi_2 - \frac{W}{\kappa } t)  }{\sqrt{g_{S^3} }}
\frac{  \delta^4( X_M - x_M) }{\sqrt{G_M}}.
\end{eqnarray}
The strength of all the stress tensor components then becomes $\frac{\Delta}{R}$, which is the expected situation for a geodesic. Compactifying on $S^3\times M$,  we obtain the following sources. The Einstein equations or the metric in $3d$  gets a source term  given by the stress tensor 
\begin{eqnarray} \label{metcirsource}
T^{tt}  = \frac{\Delta}{  R}    \frac{ \delta(r)\delta( \phi) }{r}.
\end{eqnarray}
Note that we have integrated the stress tensor in 10 dimensions given in (\ref{circular10d}) in all the compact directions. The angle $\phi_1$ needs to be integrated from $0$ to $2\pi m $, since the circular string is wrapped $m$ times along $\phi_1$. The metric fluctuations with one of the components in  $\phi_1$ or $\phi_2$ results in $2$ 
$U(1)$, Chern-Simons  gauge fields $A_\mu^{\phi_1}, A_\mu^{\phi_2}$ respectively. They are sourced by the following current density in $3d$
\begin{eqnarray}
J^t_{\phi_1}  = \frac{ (J + 2 k qm) }{ R}   \frac{ \delta(r)\delta( \phi) }{r} , \qquad 
J^t_{\phi_2} = \frac{J}{R}  \frac{ \delta(r)\delta( \phi) }{r} .
\end{eqnarray}
Finally there are three scalars  in $3d$  labelled as $\psi_1, \psi_2, \psi_3$ which are sourced by 
\begin{eqnarray}
J_{\psi_1} &=&  \frac{ ( J +  2 k qm )^2  -k^2 m^2}{ R \Delta }  \frac{ \delta(r)\delta( \phi) }{r} , \\ \nonumber
J_{\psi_2} &=&  \frac{ J^2 -k^2 m^2  }{ R\Delta }  \frac{ \delta(r)\delta( \phi) }{r}, \\ \nonumber
J_{\psi_3} &=&  \frac{J  (J +  2 k qm ) - k^2 m^2 } { R\Delta }  \frac{ \delta(r)\delta( \phi) }{r}.
\end{eqnarray}
At least one of the scalars is massless since, as we have argued earlier in section \ref{seccftdualads3}, the dilaton couples to the scalar components of the stress tensor.  We would have to study the compactification in more detail to figure out the mass of the other scalars. Observe that we can take the winding number $m=0$ and use $\Delta = J$ to obtain the sources in the compactified theory for the case of the circular string without world sheet spatial dependence. We can now proceed to find the backreacted solution just as we have done for the geodesic in 
section  \ref{seccftdualads3}. The difference would be that there is an additional gauge field, and 2 additional scalars in the bulk, which have a non-trivial solution. This implies that the operator in the CFT  ${\cal O}$ has an additional R-charge and there are $2$ additional operators with non-trivial $3$ point functions with ${\cal O}$. Finally, we can perform the coordinate transformation  in 
(\ref{coordinate transformation for boosted in-falling geodesic})  to the back reacted solution to the in-falling circular string in BTZ and demonstrate that these are dual to quenches that carry $R$ charge and expectation values of scalars. We will not carry out this analysis, but the same procedure done for the geodesic in section \ref{quenchesbtz} can be repeated here. This will result in the demonstration that the in-falling circular strings are quenches in the thermal CFT. It is clear that the backreacted metric is identical to that of the in-falling geodesic, since comparing 
(\ref{sources}) and (\ref{metcirsource}) we see that the source for the metric corrections is identical. Finally, we remark that the backreacted solution for the giant magnon solution can be treated following the same approach as for the circular string.  It is clear from our analysis of the circular string and that of the geodesic, the in-falling Magnon in the BTZ geometry will be a quench sourcing energy density, R-charges as well 
as certain scalar expectation values in the thermal CFT.

\section{Conclusions} \label{sec5}
 
 We have studied the spectrum of string states that fall into the horizon of the $BTZ \times S^3 \times M$ geometry and their dual description in the thermal field theory. The states consist of geodesics,  circular strings, giant magnons, and plane waves. These solutions were obtained by utilising the map which relates the time-like geodesic at the origin of $AdS_3\times S^3 \times M $  and spinning on $S^3$ to the in-falling particle in the BTZ geometry. This map was used to zoom into a geometry in the BTZ  black hole for which the plane wave spectrum is identical to that of $AdS_3$. We have also shown that the dual description of the classical solutions in-falling into the BTZ horizon are generalised local quenches in the thermal CFT. These quenches are not symmetric on the left and right light cones when the classical solution has velocity along the boundary of the BTZ background; they carry R-charges and expectation values of other operators in the theory.
 
 Early studies of strings in BTZ  explored the fact that it  is an orbifold of $AdS_3$ and derived properties of the spectrum with an emphasis on twisted states and integrability 
\cite{Natsuume:1996ij,Hemming:2001we,Troost:2002wk,Hemming:2002kd,Rangamani:2007fz,Berkooz:2007fe,David:2011iy,David:2012aq,Mertens:2015ola,Ashok:2021ffx,Nippanikar:2021skr,Ashok:2022vdz}.
It will be interesting to re-examine these works in the light of the fact that there are states that 
fall into the horizon of the BTZ and identify these states from the general analysis. 

The analysis done in this paper can be generalised for hyperbolic black holes in higher dimensions. This is because there is a map that takes the $AdS_d$ with $d>3$ to the hyperbolic black hole or $AdS_d$ Rindler
\cite{Casini:2011kv}. Therefore, circular strings and giant magnon solutions in $AdS_5\times S^5$ or $AdS_4\times S^7$ can be related to solutions falling into the $AdS_d$ Rindler space time. Following our analysis further would lead to a dual description of these solutions in the thermal CFT on the hyperbolic cylinder $S^1\times H_{d-2}$. It would be interesting to make this more explicit and explore other possible deformations 
of the map (\ref{coordinate transformation for boosted in-falling geodesic}) as we have done for the case of $AdS_3$. 

A by-product of our study was the observation of more general local quenches in thermal CFT. The important aspect of these quenches is that the left and right pulses on the light cone are not 
symmetric. They also carry R-charges and expectation values of other operators. It will be interesting to study these quenches in more detail, especially the evolution of entanglement entropy on the lines of \cite{David:2016pzn}. Since these quenches carry R-charges, it is possible that notions of symmetry resolved entanglement \cite{Zhao:2020qmn} might play a role.

\acknowledgments

We thank Rajesh Gopakumar for discussions at various stages of this project, whose questions led us to study the
spacetime charges of the in-falling states in BTZ. RM would like to thank Ritwick Kumar Ghosh for stimulating discussions. J.R.D thanks  S. Prem Kumar and Suvajit Majumder for an enjoyable collaboration during the difficult Covid times in which asymmetric quenches were first found in the BTZ background. RM acknowledge support of the Department of Atomic Energy, Government of India, under project no. RTI4019.

\appendix

\section{Penrose limit of Schwarzschild black hole in \texorpdfstring{$AdS_5 \times S^5$}{}}
\label{A:penrose limit of AdS5S5}
Consider the Schwarzschild black hole in $ AdS_5 \times S_5$
, the metric is given in
\begin{equation}\label{AdS5S5BhMetricAp}
    R^{-2}ds^2 = - \lr{r^2 +1- \frac{r_0^2}{r^2}}dt^2 + \frac{dr^2}{\lr{r^2 +1- \frac{r_0^2}{r^2}}} + r^2 \lr{d\phi^2 + \sin^2{\phi}\ d\Omega^2} + d\psi^2 + sin^2\psi \ d\Omega^2_4.
\end{equation}
Consider the case of a null geodesic carrying angular momentum along $\phi$, one of the circles in
the $AdS_5$ directions. The trajectory of this geodesic is given in equation \eqref{schbhgeo}.
We now perform the usual steps to take the Penrose limit along this geodesic. We first
make the change of coordinates from 
 $ (t,r,\phi)$ to $(\lambda,\xi,\theta)$ such that 
\begin{equation}\label{Plane wave condition}
    g_{\lambda\lambda} = g_{\lambda\theta} = 0 , \quad \text{and} \quad g_{\lambda \xi}= 1.
\end{equation}
In this new coordinate system, $\lambda$ is the usual affine parameter that we have used before to write down the geodesic, and $\xi$ and $\psi$ are the integration constants. We choose the integration constant in such a way that the above conditions, i.e., equation \eqref{Plane wave condition}, are met. So, one of the ways to choose the new coordinates is the following
\begin{equation}
    t(\lambda,\xi,\theta) = \int_{0}^{\lambda} \dot{t}\ d\lambda' -\frac{\xi}{E  } + J\theta \ ,\quad r(\lambda,\xi,\theta)= \int_{0}^{\lambda} \dot{r} \ d\lambda' \ , \quad \phi(\lambda,\xi,\theta) = E\theta + \int_{0}^{\lambda}\dot{\phi}d\lambda'.
\end{equation}
We substitute this transformation in \ref{AdS5S5BhMetricAp} and get 
\begin{align*}
    R^{-2}ds^2 &= \lr{-a_0 \dot{t}^2 + a_0^{-1}\dot{r}^2 + r^2\dot{\phi}^2}d\lambda ^2 - a_0E^{-2}d\xi^2 + \lr{r^2E^2-a_0J^2 }d\theta^2 + (a_0E^{-1}\dot{t})2d\lambda d\xi  \\ &+ \lr{-a_0J \dot{t} + JE}2d\lambda d\theta + (a_0JE^{-1}) 2d\xi d\theta + r^2\sin^2{\phi}\ d\Omega^2_2 + ds^2(S^5),  \\
    &= 2d\lambda  d\xi  -  a_0E^{-2}d\xi^2 + \lr{r^2E^2-J^2a_0}d\theta^2 + J a_0E^{-1}2d\xi d\theta + r^2 \sin^2{\phi}.\ d\Omega^2_2 + ds^2(S^5)
\end{align*}
Here we have redefined:  $ a_0\equiv\lr{r^2 +1- \frac{r_0^2}{r^2}}$. Now, we perform the following rescaling,
\begin{align}
     \lambda \rightarrow \lambda , \quad \xi\rightarrow \frac{\xi}{R^2},\quad \theta\rightarrow \frac{\theta}{R},\quad \psi \rightarrow \frac{\psi}{R},
\end{align}
and expanding the metric in the large R limit, we obtain
\begin{align}\label{AdS5.S5 BH plane wave metric in Rosen}
    ds^2 &= 2d\lambda d\xi + \lr{r^2E^2-J^2a_0}d\theta^2 + r^2\sin^2\lr{\int_{0}^{\lambda}\frac{J}{r^2}d\lambda'}\ ds^2(\mathbb{R}^2) + ds^2(\mathbb{R}^5),  \\
    &= 2d\lambda d\xi +  {\sum\limits_{i=1}^8 \lr{C_{i} dy^i}^2 } . 
\end{align}
The above metric, equation \eqref{AdS5.S5 BH plane wave metric in Rosen}, is the plane wave metric in the Rosen coordinate.  We go to Brinkmann coordinates by the following coordinate transformation
\begin{equation}\label{Brinmann coordinate transformationa}
    \lambda=x^+ ,\quad \xi=x^- +\frac{1}{2}\sum\limits_{i} \frac{1}{C_i}\lr{\frac{\partial C_i}{\partial x^+}}x^ix^i,\quad y^i= C^{-1}_ix^i.
\end{equation}
Then the metric becomes 
\begin{align}
    ds^2&= 2dx^+dx^- + \lrt{\sum\limits_{i} {\frac{C^{''}(x^+)}{C_i} x^ix^i}}\lr{dx^+}^2 + \sum\limits_{i}dx^idx^i, \nonumber\\
    &=2dx^+dx^- + \sum\limits_{i}  \lr{ M_{ii}x^ix^i}\lr{dx^+}^2 + \sum\limits_{i}dx^idx^i\nonumber.
\end{align}
To evaluate the matrix M, we first write the derivative with respect to $\lambda$ in terms of
derivative with respect to r as,
\begin{align}\label{derivative w.r.t u in terms of r}
    \frac{d^2}{d\lambda^2}= \frac{d}{d\lambda}\lr{\dot{r}\frac{d}{dr}}= \dot{r}^2\frac{d^2}{dr^2} + \dot{r}\lr{\frac{d\dot{r}}{dr}}\frac{d}{dr}.
\end{align}
Using the above relation, we compute all the components of the M matrix 
\begin{align}
    M_{11}&= r^{-1}\frac{d}{dr}\lrt{\dot{r}\frac{d}{dr}\lr{\dot{r}r}}= \frac{4J^2r_0^2}{r^6};\quad M_{44}=...=M_{88}=0,\nonumber\\
    M_{22}&=M_{33}= \frac{\dot{r}}{r\ \sin{\lr{\int \frac{J}{r^2}du}}}\frac{d}{dr}\lrt{ \dot{r}\frac{d}{dr}\lrs{r\sin{\lr{\int \frac{J}{r^2}du}}}}=-\frac{2J^2r_0^2}{r^6}.
\end{align}
This results in the following pp wave metric in Brinkmann coordinates
\begin{align}
    ds^2 = 2dx^{+}dx^- + \lrt{\frac{4J^2r_0^2}{r^6}x_1^2 -\frac{2J^2r_0^2}{r^6} (x_2^2 +x_3^2)}\lr{dx^+}^2 + \sum\limits_{i=1}^{3}dx_i^2 + \sum\limits_{i=1}^{5}dy_i^2.
\end{align}
Now we examine the situation in which the null geodesic has angular momenta along one of the circles of $S^5$. 
This case was addressed in \cite{PandoZayas:2002dso} 
with a slightly different metric for the $AdS_5$ Schwarzschild black hole. 
The trajectory of this geodesic is already given by the equation \eqref{geoschspher}. As before, we follow similar steps to take the Penrose Limit. We begin with a change of coordinate from ($t,r,\psi$) to ($\lambda,\xi,\chi$) which is given by
\begin{align}\label{cordChngAds5S5Bh}
t(\lambda,\xi,\chi) = \int_{0}^{\lambda} \dot{t}\ d\lambda' -\frac{\xi}{E  } + J\chi \ ,\quad r(\lambda,\xi,\chi)= \int_{0}^{\lambda} \dot{r} \ d\lambda' \ , \quad \psi(\lambda,\xi,\chi) = E \chi + J\lambda .
\end{align}
Substituting this coordinate transformation in the metric \eqref{AdS5S5BhMetricAp}, we perform the following rescaling
\begin{align}
     \lambda \rightarrow \lambda , \quad \xi\rightarrow \frac{\xi}{R^2},\quad \chi\rightarrow \frac{\chi}{R},\quad \phi \rightarrow \frac{\phi}{R}.
\end{align}
Now expanding the metric in the large R limit, we obtain the pp wave in the Rosen coordinates
\begin{align}
    ds^2 &= 2d\lambda d\xi + \lr{E^2-J^2a_0}d\theta^2 + r^2 ds^2(\mathbb{R}^3) +\sin^2{(J\lambda)} \ ds^2(\mathbb{R}^4).
\end{align}
Next, we go to the Brinkmann coordinates and calculate the mass matrix
\begin{align}
    \tilde{M}_{11} &= \lrt{{E^2-J^2\lr{r^2 +1- \frac{r_0^2}{r^2}}}}^{-\frac{1}{2}}\frac{d^2}{d\lambda^2}\lr{\sqrt{1-J^2\lr{r^2 +1- \frac{r_0^2}{r^2}}}}, \nonumber\\
    &=\frac{d}{dr}\lr{\dot{r}\frac{d}{dr}(\dot{r})}
     = -J^2\lr{1-\frac{3 r_0^4}{r^4}},\\
    \tilde{M}_{22} &= \frac{1}{r}\frac{d^2}{d\lambda^2}(r) = \frac{1}{r}\lrt{\dot{r}^2\frac{d^2}{dr^2} + \dot{r}\lr{\frac{d\dot{r}}{dr}}\frac{d}{dr}} (r) =\dot{r}\lr{\frac{d\dot{r}}{dr}}^2 = -J^2\lr{1+\frac{ r_0^4}{r^4}} = \tilde{M}_{33}=\tilde{M}_{44},\\
    \tilde{M}_{55} &= \frac{1}{\sin{(J\lambda)}}\frac{d^2}{d\lambda^2}\lr{\sin{(J\lambda)}} = -J^2 = \tilde{M}_{66}=\tilde{M}_{77}=\tilde{M}_{88}.
\end{align}
Finally, we obtain
 \begin{equation}
    ds^2 = 2 dx^+dx^- - J^2\lrt{\lr{1-\frac{3r_0^4}{r^4}} x_1^2 + \lr{1 + \frac{r_0^4}{r^4}} z_3^2 +z_4^2}\lr{dx^+}^2 +dz_3^2+ dz_4^2,
\end{equation}
where $z_3$ parametrizes $\mathbb{R}^3$ and $z_4$ parametrizes $\mathbb{R}^4$. This solution is supported by five-form flux, which is given by
\begin{equation}
    F_5 = R^{4}\lrt{\text{vol}(AdS_5) + \text{vol}(S^5)},
\end{equation}
where
\begin{align}
    \text{Vol}(AdS_5)=r^3 dt\wedge dr\wedge \text{Vol}(S^3),\quad \text{Vol}(S^5)=\sin^4\psi \ d\psi\wedge\text{Vol}(S^4).
\end{align}
Substituting the coordinates transformation given in \eqref{cordChngAds5S5Bh} in the above equation and taking the large R limit, we obtain
\begin{align}
   \tx{Vol}(AdS_5)&= R^4 \ r^3 dt\wedge dr\wedge \tx{Vol}(S^3) ,\nonumber\\
    &= R^4 \ r^3 \lr{t_\lambda d\lambda - \frac{d\xi}{E} + J d\phi}\wedge\lr{ r_\lambda d\lambda}\wedge \tx{Vol}(S^3) ,\nonumber\\
    &= R^4 \ r^3 r_\lambda \lrt{-\frac{d\xi}{E} \wedge d\lambda \wedge \tx{Vol}(S^3) + J \ d\phi\wedge d\lambda \wedge \tx{Vol}(S^3)  },\nonumber\\   
    &= J \  d\phi \wedge (r_\lambda d\lambda) \wedge (r^3\tx{Vol}(\mathbb{R}^3)),\\
     \tx{Vol}(S^5)&= R^4 \ \sin^4{\psi} \ d\psi \wedge \text{Vol}(S^4),\nonumber\\
    &= R^4 \ \sin^4{(J \lambda +E \phi)} \lrs{ J d\lambda + E d\phi}\wedge  \text{Vol}(S^4), \nonumber \\
    &=  J\ d\lambda \wedge \sin^4{(J \lambda)}  \text{Vol}(\mathbb{R}^4).\nonumber
\end{align}
Going to the Brinkmann coordinates, we obtain
\begin{equation}
    F_5 = J \ dx^+ \wedge\lr{ dx_1 \wedge dx_2 \wedge dx_3 \wedge dx_4  + dx_5\wedge dx_6\wedge dx_7\wedge dx_8}.
\end{equation}

\bibliographystyle{JHEP}
 \bibliography{biblio.bib}

\end{document}